# The MEVIR 2 Framework: A Virtue-Informed Moral–Epistemic Model of Human Trust Decisions

Daniel Schwabe
 dschwabe@gmail.com

## Abstract

The MEVIR 2 framework (an improved version of the original MEVIR framework) addresses a critical challenge in our modern information environment: understanding how real people actually make trust decisions amid complexity, polarization, and misinformation. Unlike classical rational-agent models that focus on ideal logical calculation, MEVIR 2 recognizes that human trust involves an intricate interplay of reasoning, character, and rapid moral intuitions. It is a descriptive and explanatory model of human trust decisions built on three interacting pillars: a procedural model of evidence elaboration; a virtue-theoretic model of the epistemic agent; a hybrid Moral Model that integrates morality as evolutionary cooperative problems – using Morality-as-Cooperation (MAC) model - with culturally shaped morality - using the Extended Moral Foundations Theory (EMFT). This explains why certain authorities, facts, and trade-offs feel trustworthy or salient to different people.

A key innovation of MEVIR 2 is its introduction of "Truth Tribes" (TTs)—stable clusters of individuals who share similar configurations across all three structural layers of the model. These are not mere social groups or ideological factions, but emergent communities with aligned procedural, virtue-theoretic, and moral-epistemic profiles. By making TTs the primary unit of analysis, the framework explains why polarization persists and how different communities construct internally coherent "trust lattices" that remain mutually unintelligible to outsiders. The model also incorporates ontological distinctions between Truth Bearers and Truth Makers, showing that disagreements often stem not just from differing beliefs about what is true, but from fundamentally different views about what aspects of reality can make propositions true.

The practical applicability of MEVIR 2 is demonstrated through detailed case studies on vaccination mandates and climate policy, revealing how different moral configurations lead people to select different authorities, evidential standards, and trust anchors—effectively constructing separate moral-epistemic worlds. Special attention is given for when agents operate within TTs.

The framework reinterprets cognitive biases as systematic failures of epistemic virtue and offers a foundation for designing decision-support systems that could enhance metacognition, make trust processes transparent, and foster more conscientious public reasoning across divided communities.

---

# 1   Introduction: Trust Decisions in a Cooperative Information Ecosystem

The contemporary information ecosystem confronts individuals with an unfamiliar challenge: how to make responsible trust decisions in an environment saturated by complexity, polarization, and strategically engineered misinformation. Classical models of the rational agent, grounded in probability and utility, describe how an ideal calculator might decide, but they do not capture how real human beings actually come to trust people, institutions, or claims. Real-world trust is not a purely logical derivation; it is a process in which reasoning, character, and rapid, pre-reflective moral intuitions are tightly intertwined. A citizen deciding whether to vaccinate a child, support a climate policy, or believe an investigative report faces not only incomplete information but also a torrent of emotionally charged narratives that frame these choices in radically different moral terms.

Rational-agent models in economics and decision theory describe agents who compile evidence, compute probabilities, and select options that maximize expected utility (Gilboa 2009). These models are useful in stylized environments, yet they struggle to explain why intelligent and well-intentioned people, when faced with essentially the same body of evidence on issues such as climate change or vaccination, arrive at incompatible and fiercely defended conclusions. The persistence of systematic cognitive biases, the power of identities and group loyalties, and the depth of moral disagreement all suggest that human trust formation follows a richer and less transparent logic than these models anticipate.

The central challenge, therefore, is not merely to note that human reasoning deviates from an ideal of deductive rationality, but to understand the architecture of those deviations. Biases and emotional reactions are not random noise; they are structured outputs of a cognitive–moral system shaped by evolutionary pressures, cultural learning, and personal history. Any model that aims to describe trust must therefore account for the interaction of multiple layers: procedural reasoning, epistemic character, and intuitive moral orientation.

In digital environments where claims are generated and propagated with minimal friction and weak curation, the question "whom or what should I trust?" becomes central. Formal and computational models of trust in information have made progress in representing chains of evidence, trust anchors, and policies for evaluating sources. They show how an agent might recursively search for supporting statements until a chain terminates in pre-accepted beliefs or authorities. Yet in these models the most important elements—those primitive beliefs, trust anchors, and evaluative policies—are treated as unexplained starting points. The models describe the syntax of justification but leave the semantics of trust largely opaque.

MEVIR 2 is an improved version of the original MEVIR framework (Schwabe 2025), extending the Moral Model to incorporate Morality as Cooperation, and introducing the concept of Truth Tribes. It is proposed as a more psychologically realistic and ontologically explicit response to this challenge. It treats trust not as a single moment of assent but as a process that unfolds at three interconnected levels. At the procedural level, the agent constructs and revises evidence structures. At the virtue level, the agent's intellectual character—courage, humility, open-mindedness—governs when and how these procedures are activated. At the moral level, a hybrid of Morality-as-Cooperation (MAC) and Extended Moral Foundations Theory (EMFT) explains how cooperative problems and intuitive moral responses shape which possibilities appear desirable or even thinkable.

Morality-as-Cooperation provides an evolutionary and game-theoretic account of morality as a set of solutions to recurring problems of cooperation such as helping kin, sustaining groups, reciprocating, coordinating hierarchies, and respecting property (Curry 2016; Curry, Mullins, and Whitehouse 2019). EMFT models morality as a set of intuitive "foundations," such as Care, Fairness, Loyalty, Authority, Liberty, and Purity, that function like taste receptors for moral content and are unevenly emphasized across individuals and political cultures (Haidt 2012). MEVIR 2 integrates these perspectives by treating MAC's cooperative problems as the deep structure of moral life, and EMFT's foundations as the intuitive interface through which these problems are experienced and acted upon.

The result is a framework in which disagreements about controversial topics such as vaccines or climate policy are not merely disagreements about facts. They are disagreements about which cooperative problems are primary, which groups deserve priority, which authorities are legitimate, what trust policies should be applied, and which harms count morally. MEVIR 2 aims to make these hidden structures explicit and thus provide a language in which trust conflicts can be described, diagnosed, and, at least in principle, addressed.

Appendix I provides an actionable summary for anyone who intends to apply MEVIR 2 to a given narrative ((e.g., a speech, article, or social media thread, proposed legislation, etc...), and gives a concrete example for the Generative AI vs Artificial General Intelligence (AGI) debate.

---

# 2  Ontological Foundations: Truth Bearers, Truth Makers, and Unpacking

Understanding trust requires greater precision about what, exactly, is being trusted. In everyday discourse people exchange sentences, slogans, and data visualizations; yet these textual or visual artifacts are only the outer layer of a deeper relation between representation and reality. The framework adopts a correspondence-theoretic view of truth (David 2025): a claim is true when there exists something *in reality* that makes it true.

Truth Bearers are the entities that can be true or false. They include propositions, claims, sentences, database entries, or model states. When someone says, "this vaccine is safe" or "this decade will be decisive for the climate," they introduce truth-bearers (the "claims") into the conversational space. Truth Makers are the aspects of reality that confer truth on these bearers, e.g., the actual pattern of adverse events in vaccinated populations, the distribution of temperature changes and extreme weather events, the causal structure of virus transmission or climate systems. In other words, if the Truth Maker can be shown to be true in "the real world", then the Truth Bearer ("the claim") is true. A truth-bearer is trustworthy only insofar as the agent believes that an appropriate truth-maker exists and that available information reliably tracks it.

Disagreement often arises not because agents read different sentences, but because they implicitly posit different admissible truth-makers for those sentences. The same text — "this vaccine is safe" — may be unpacked ontologically as a claim about long-term statistical outcomes in a population, about the absence of contaminants that violate a purity norm, or about the integrity of the institutions that licensed the vaccine. If two agents silently use different kinds of truth-makers, they may appear to talk about the same claim while in practice addressing different realities.

Ontological Unpacking is the method by which the implicit commitments of a truth-bearer are made explicit (Guizzardi and Guarino 2024, Amaral 2025). To unpack a claim is to ask what kinds of objects, events, relations, and dispositions must exist for the claim to be true. When the claim is "the virus is airborne," unpacking reveals a structure that includes viral particles, biological hosts, modes of transmission, spatial relations, and normative thresholds for what counts as "airborne" in epidemiological practice. The precision gained through unpacking is essential for understanding how different agents can sincerely disagree about "the same" sentence because they are tracking different slices of reality.

## 1.1. Unpacking in Public Health

The domain of public health illustrates how compressed truth-bearers can hide complex ontological structures. Surveillance databases frequently store viral information as strings of characters representing genetic sequences. An ontological model such as the Viral Conceptual Model makes explicit that a virus is an entity with dispositions such as infectivity and virulence; an infection is an event in which such dispositions manifest; and a mutation is not merely a textual change, but an event in the evolutionary history of a viral lineage (Bernasconi et al. 2022; Guizzardi et al. 2021).

When public health agencies announce that a new variant is "of concern" because it is "more transmissible," the truth-bearer seems simple. Yet citizens may unpack this statement differently. For some, the relevant truth-maker is an experimentally demonstrated increase in viral load or basic reproduction number. For others, shaped by Purity and Liberty intuitions, the relevant truth-maker is whether the variant is associated with interventions that expand state authority or alter perceived bodily integrity—lockdowns, mask mandates, vaccine passports. If the latter are treated as the real locus of concern, then scientific data about viral dynamics may feel orthogonal to what is morally at stake.

## 1.2. Unpacking in Climate Change

The concept of "climate risk" offers a parallel example. In expert discourses, risk is not a floating label but an ontologically structured compound: it involves a threat event, a vulnerability, and an impact defined relative to a goal (Adamo et al. 2024). For climate scientists and environmentalists, the relevant truth-makers for "high climate risk" are trajectories that lead to ecological tipping points, mass species loss, and severe harms for vulnerable communities. For economic stakeholders whose primary goal is sustained growth or national competitiveness, the salient truth-makers may instead be projected GDP losses from aggressive mitigation policies.

Two agents may agree that "risk is high" but disagree about whether this is acceptable or urgent, because they tacitly prioritize different goals and thus different impacts. Once again, trust in a climate claim is inseparable from an ontological judgment about which risks matter, to whom, and in which cooperative frame—local industries, national prosperity, global justice, or intergenerational stewardship.

MEVIR 2 sets the stage for its three-pillar architecture by bringing truth-makers and unpacking to the foreground. The procedural pillar describes how claims are recursively linked to supporting statements. The virtue pillar describes how agents manage that process in a conscientious or negligent way. The hybrid moral pillar describes how cooperative concerns and intuitive foundations determine which unpacked realities are considered morally salient in the first place.

# 3   The Constituent Architectures of MEVIR 2

MEVIR 2 rests on three interlocking architectures. The first is a procedural model of trust elaboration that captures the mechanical structure of evidence gathering. The second is a virtue model that characterizes the epistemic agent in the process of building the structure of evidence. The third is a hybrid MAC–EMFT moral model that supplies the intuitive and cooperative background against which trust decisions acquire meaning.

## 1.3. The Procedural Model: Mechanics of Trust Elaboration

At the procedural level, "trust in information" is defined functionally. An agent trusts a claim when that claim is used to guide or justify an action in each context. When someone decides to vaccinate a child, oppose a climate tax, or accept a medical treatment, the underlying decision is mediated by a network of supporting statements about safety, efficacy, economic consequences, and moral obligations. The procedural model describes how such networks are built and revised.

When confronted with a claim whose truth value is uncertain or that conflicts with existing beliefs, the agent initiates an elaboration process. This process consists of a recursive search for evidential statements whose acceptance would support or undermine the original claim. The agent assesses each candidate piece of evidence in turn, possibly triggering further elaboration. Over time this yields a structure of interdependent statements: a trust chain if the structure is linear, or more realistically a trust lattice where different claims share common sub-chains (i.e., a statement may support or refute several other statements at the same time).

The recursive search for evidence cannot continue indefinitely. The procedural model specifies that the process must terminate when it reaches a *trust anchor*. An anchor is a statement that requires no further justification for the agent, *within the current context*. The model identifies five conditions under which the recursion stops:

1. Pre-trusted Statements: Evidence established through a prior process.
2. Beliefs: Statements accepted unconditionally.
3. Accepted Authorities: The statement is made by a source accepted under a prevailing social norm, such as scientific institutions, religious leaders, or trusted peers.
4. Evidence Exhaustion: No further info is available.
5. Resource Exhaustion: Time, computational resources, or cognitive capacity runs out.

Of these, "Beliefs" and "Accepted Authorities" are the most analytically significant. They represent the internal and external endpoints of justification. In MEVIR 2.0, these anchors are not random; they are fundamentally determined by the Moral Model.

In practice, anchors do not directly access truth-makers, but information proxies: a certificate for a birth event; a laboratory certification stands in for a complex experimental reality; a legal certificate for a historical event; or a scientific consensus for a vast literature.

Once a lattice has been constructed, the agent applies trust policies to decide which claims will be treated as factual for purposes of the contemplated action. Trust policies include criteria about source reliability, standards of evidence, and inferential rules. These policies may be explicit but are often tacit habits: a physician may habitually weight randomized controlled trials more heavily than anecdotal reports; a layperson may discount any source associated with a distrusted political faction.

The procedural architecture also incorporates belief revision. When new, trusted information enters the system and conflicts with existing claims, the agent must restore consistency. The principle of minimal change suggests that agents will seek revisions that alter their belief set as little as possible while accommodating the new statement (Hunter and Booth 2015). However, MEVIR 2 allows for a non-destructive approach in which previous lattices are not erased but held in reserve, recognizing that new evidence may later rehabilitate them (Pereira, Tettamanzi, and Villata 2011). This explains the "stickiness" of misinformation; if accepting a correction requires destroying a central "Belief" anchor or discrediting a central "Authority," the computational cost to the lattice is too high, and the agent will likely reject the correction instead.

In isolation, the procedural model remains agnostic about why certain statements function as anchors or why particular policies are applied. For this, MEVIR 2 turns to its virtue and moral pillars.

## 1.4.  The Virtue Model: The Conscientious Epistemic Agent

The virtue pillar of MEVIR 2 draws on Linda Zagzebski's neo-Aristotelian virtue epistemology (Zagzebski 1996). Rather than focusing solely on properties of beliefs—justification, warrant, reliability—it centers the epistemic agent and the character traits that make them a good or bad knower or "knowledge producer".

The virtue pillar interacts with the ontological pillar: a virtuous agent is not only conscientious in gathering evidence but also in clarifying the relevant truth-makers and admissible proxies for a given context. Correctly distinguishing between biological and legal truth-makers, for instance, is part of what it means to reason appropriately about inheritance or public health.

The point of departure is the conscientious agent: someone motivated by a genuine desire to know the truth and avoid error. This motivation grounds epistemic self-trust, the basic confidence that one's own cognitive faculties—perception, memory, reasoning—are generally reliable unless specific defeaters are present. Self-trust is not an optional add-on; it is the condition of possibility for any rational inquiry. Without it, skepticism would paralyze action.

Zagzebski models belief formation as unfolding along two paths. In situations where the agent judges themselves to be sufficiently competent, they adopt a path of direct reliance: they use their own faculties to evaluate evidence and reach a conclusion (Path 1). In situations where the agent recognizes their limits, they defer to epistemic authorities (Path 2). Deference is not blind faith but itself a structured act of judgment: the agent must identify who counts as a genuine authority in a specific domain and distinguish them from charlatans or propagandists. Even deference is therefore rooted in self-trust; the agent must trust their own ability to track expertise.

Herbert Simon's notion of bounded rationality provides the practical rationale for this two-path architecture (Simon H 1990). Human agents lack the time, computational power, and specialized knowledge to personally evaluate all relevant evidence for every complex decision. In many domains, especially technical ones like virology or climate modeling, an agent's only realistic option is to seek out and evaluate authorities. The virtue of epistemic humility is therefore not a weakness but a rational adaptation to one's own recognized cognitive limits.

Intellectual virtues such as open-mindedness, intellectual courage, attentiveness, and intellectual perseverance govern how agents navigate these paths. They influence when direct evaluation is attempted, when deference is sought, how counter-evidence is handled, and how biases are corrected. Conversely, epistemic vices—closed-mindedness, arrogance, laziness, and cowardice—systematically distort the use of the procedural machinery, which often manifests itself as biases.

In MEVIR 2, many familiar cognitive biases are reconceived as failures of virtue, emerging not as isolated quirks but as systematic failures in how we navigate the fundamental tension between trusting ourselves and trusting others. These biases, then, are not random glitches but systematic patterns emerging from how we mismanage the choice between self-trust and deference to others. We remain in Path 1 when we should switch to Path 2. We corrupt Path 2 by selecting validators rather than experts. We contaminate Path 1 with skewed evidence or arbitrary anchors. Understanding these patterns means understanding not just isolated mistakes but the deeper architecture of how we—and how we fail to—navigate the uncertain terrain between our own limited knowledge and the vast expertise we cannot personally possess.

Consider *overconfidence* bias and its cousin, the *Dunning-Kruger* effect. Here we witness a classic failure of epistemic humility—the inability to recognize the limits of one's own knowledge. The amateur investor, flushed with success from a handful of lucky trades, becomes convinced of their market mastery. They should recognize their limitations and seek expert guidance, switching from Path 1 to Path 2, but instead they remain stubbornly trapped in self-reliance, making increasingly reckless bets while dismissing the wisdom of professional analysts.

*Confirmation bias* represents a different kind of failure—a corruption of the very mechanism meant to protect us from our own limitations. When we should be seeking the most reliable authorities to guide us, we instead hunt for voices that validate what we already believe. The person convinced of a political candidate's corruption doesn't genuinely defer to expertise; they curate a media diet that transforms Path 2 into an echo chamber, selecting (pseudo) experts who amplify rather than challenge their intuitions.

The *availability heuristic* corrupts judgment from within. When dramatic shark attack footage dominates someone's memory, they mistake vividness for probability. Their direct evaluation—Path 1—operates on contaminated data, a skewed sample that feels representative but isn't. The beach vacation gets cancelled not because of rational risk assessment but because emotionally charged memories have hijacked the evidence base.

*Anchoring bias* reveals how easily our supposedly independent judgment can be manipulated. That first number in a salary negotiation—whether reasonable or absurd—becomes a gravitational force around which all subsequent thinking orbits. The person believes they're exercising direct judgment, but their Path 1 evaluation has been contaminated from the start, operating within an artificially constrained frame they didn't choose.

The *bandwagon effect* shows us mistaking popularity for truth. Rather than deferring to genuine expertise, people outsource their beliefs to the crowd, prioritizing social comfort over accuracy. The individual adopting a trendy opinion or fashion has technically engaged Path 2—they're following others—but they've confused social consensus with reliable authority, choosing conformity over competence.

The *fundamental attribution* error demonstrates overconfidence in our intuitive judgments about others. Seeing a coworker arrive late, we rush to conclusions about their character—lazy, disorganized—without recognizing how little we actually know. We should maintain epistemic humility, acknowledging that situational factors might explain the behavior, but instead we trust our snap judgments, remaining locked in an inappropriately confident Path 1 evaluation.

*Reactance* represents perhaps the most perverse failure: deliberately choosing error to preserve autonomy. The teenager told to avoid certain friends seeks them out specifically because they were warned away. Here Path 2 becomes inverted—authority triggers automatic rejection rather than consideration. The emotional need for independence overpowers the epistemic goal of finding truth, turning expert guidance into a compass pointing the opposite direction.

The *halo effect* shows how we extend authority beyond its proper domain. A successful business leader's competence in commerce creates an illusion of general wisdom. Voters trust their opinions on foreign policy and public health, not because they've evaluated expertise in those areas, but because success in one domain has cast a glow over everything else. This also happens with modern-day celebrities and influencers. Path 2 gets corrupted when we fail to recognize that authority is domain-specific, not universal.

Finally, the *false consensus effect* reveals how our limited perspective can masquerade as universal truth. The heavy social media user, surrounded by similar users, genuinely believes their platform dominates society. They're stuck in Path 1, treating their narrow experience as a reliable sample of the whole, never recognizing the need to look beyond themselves. Their direct evaluation feels certain because they mistake their bubble for the world.

Even the most conscientious agent does not operate in a vacuum. The choices about where to apply effort, which alternatives to consider, and which authorities to treat as salient are shaped by underlying moral concerns. These concerns are the focus of the hybrid MAC–EMFT moral pillar.

## 1.5.  The Moral Model: A Hybrid Morality-as-Cooperation / Extended Moral Foundations Architecture

The moral pillar in MEVIR 2 aims to explain the pre-rational and intuitive forces that shape trust decisions. It does so by integrating Morality-as-Cooperation (MAC) with Extended Moral Foundations Theory (EMFT). MAC provides a first-principles account of the problems that morality solves; EMFT provides an empirically grounded account of the intuitive patterns through which people experience and respond to those problems.

### 3.3.1  Morality-as-Cooperation: Moral Problems as Cooperative Games

MAC begins from the premise that morality is about cooperation. Over evolutionary time, individuals and groups that found stable ways of cooperating tended to fare better than those that did not; morality is thus best understood as a set of strategies for solving recurring cooperation problems under different conditions (Curry 2016; Curry, Mullins, and Whitehouse 2019).

MAC identifies a set of core cooperative problems and associated moral domains.

*Kin* cooperation generates norms that demand special care for family members and close relatives.

*Group* cooperation demands norms of loyalty and sacrifice that hold coalitions together.

*Reciprocity* addresses the problem of repeated interactions among non-kin, giving rise to expectations of returning benefits and punishing cheaters.

*Heroism* involves the willingness to accept personal risk for collective gain, as in warfare or emergency response.

*Deference* concerns the efficient organization of hierarchies by coordinating leaders and followers.

*Fairness* addresses conflicts over divisible resources, encouraging proportional or equal allocations to prevent destructive disputes.

*Property* norms solve the problem of who may use what, under which conditions, thereby reducing costly conflicts over resources.

In MAC, these domains are not arbitrary. Each corresponds to a class of games in evolutionary game theory: kin selection games, coordination games, repeated prisoner's dilemmas, hawk–dove conflicts, and so on. Moral norms are therefore seen as stable strategies that emerged because they facilitated cooperation and reduced the costs of conflict in ancestral environments.

### 3.3.2 Extended Moral Foundations: Intuitive Moral "Taste Buds"

EMFT approaches morality from the psychological side, treating it as a system of intuitive responses that are fast, automatic, and often affective (Haidt 2012). It proposes that human beings possess a set of moral "foundations" that are triggered by particular patterns of social information. These foundations function like taste receptors: they provide an immediate sense that something is right or wrong before explicit reasoning begins.

The extended version of the theory distinguishes at least six such foundations.

*Care/Harm* tracks suffering and nurturance;

*Fairness* tracks justice as equity or proportionality;

*Liberty* tracks resistance to domination;

*Loyalty* tracks allegiance to in-groups;

*Authority* tracks respect for legitimate hierarchies;

*Purity* tracks concerns about contamination and sanctity.

In addition, empirical and theoretical work has distinguished two facets within Fairness: fairness-as-equity, which emphasizes equal treatment and outcomes, and fairness-as-proportionality, which emphasizes rewards in proportion to contribution (merit-based).

EMFT research has shown that these foundations are not weighted equally across individuals and cultures. Political liberals tend to emphasize Care, Liberty, and fairness-as-equity, while conservatives tend to distribute weight more evenly across all foundations and give greater salience to Loyalty, Authority, Purity and fairness-as-proportionality. This asymmetry helps explain why people with different moral profiles can look at the same policy and experience it through very different moral lenses.

### 3.3.3 Mapping MAC Domains to EMFT Foundations

MEVIR 2 treats MAC and EMFT as complementary rather than competing models. MAC identifies the problems that moral systems must solve; EMFT describes the intuitive responses through which individuals participate in those solutions. The mapping between them is many-to-many, but several robust patterns can be gleaned.

**Kin Selection (The Family Game)**

Genes naturally flow through family lines, creating an evolutionary puzzle: how can individuals maximize the propagation of their genetic material? Hamilton's Rule provides the answer—by investing resources in relatives who share portions of one's genome, an individual indirectly promotes their own genetic legacy.

This biological imperative manifests psychologically through the Care/Harm foundation in Extended Moral Foundations Theory. The tender feelings that compel us to protect vulnerable children, nurture our offspring, and extend compassion to those in need are not abstract virtues but rather the emotional machinery implementing kin selection at the cognitive level.

When it comes to trust, this evolutionary game creates a powerful bias: agents are hardwired to extend high trust toward those labeled as "Family." This trust anchor operates so strongly that even fictive kinship—political rhetoric employing terms like "Brothers and Sisters"—can activate these ancient circuits and grant speakers a privileged status in our trust hierarchies, regardless of actual genetic relationship.

**Mutualism (The Group Game)**

Throughout evolutionary history, humans discovered that collective action unlocked possibilities unavailable to individuals—hunting megafauna, defending territory, building structures—but only if groups could maintain cohesion and reliably distinguish members from outsiders. The mutualism game rewards coordination while punishing defection, creating selection pressure for psychological mechanisms that bind individuals to their coalitions.

This pressure gave rise to the Loyalty/Betrayal foundation, the suite of emotions we experience as tribal allegiance, patriotic fervor, and the sharp psychological boundary between "us" and "them." These feelings are not cultural accidents but evolved solutions to the challenge of maintaining group cooperation.

The trust implications run deep: when evaluating truth claims, information originating from the ingroup passes through a "Loyalty" filter that asks not merely "Is this accurate?" but "Is this loyal?" Dissent from fellow group members registers not as honest disagreement or factual correction but as betrayal—a defection from the mutualism game that threatens the entire cooperative structure, triggering emotional responses far more intense than mere intellectual disagreement would warrant.

## Social Exchange (The Reciprocity Game)

The logic of reciprocal cooperation—"I scratch your back, you scratch mine"—offers tremendous benefits but harbors an inherent vulnerability: the temptation to accept favors while refusing to reciprocate, the classic Prisoner's Dilemma that threatens all cooperative ventures. Evolution equipped humans with sophisticated psychological tools to navigate this treacherous landscape, crystallized in the Fairness/Cheating foundation of moral psychology.

Interestingly, this foundation manifests in two distinct forms that map onto political orientations: an Equity emphasis more common among left-leaning individuals, and a Proportionality emphasis favored by those on the right, both representing different strategies for maintaining the ledger of social exchange.

The trust implications of this game are profound and unforgiving: trust fundamentally requires credible signals of reciprocity, which is why hypocrisy—publicly signaling adherence to one set of rules while privately following another—proves so devastating to reputations. Hypocrisy doesn't merely indicate inconsistency; it identifies the source as a "Defector" or "Free-rider" in the exchange game, someone who exploits the cooperative structure while refusing to contribute their fair **share**, justifying the immediate withdrawal of trust that such revelations typically trigger.

## Conflict Resolution: Hawk/Dove (The Contest Game)

When resources prove scarce and multiple agents desire the same valuable item, direct violent conflict becomes a possibility—but violence carries enormous costs in injury, death, and squandered energy. Game theory identifies two primary strategies in such contests: the "Hawk" approach of aggressive fighting, and the "Dove" strategy of retreat or submission, with the optimal solution depending on what others in the population are doing.

Human psychology evolved two complementary moral foundations to navigate these contests without constant bloodshed. The Authority/Subversion foundation implements the "Dove" strategy psychologically, creating the capacity to recognize legitimate rank, defer to established hierarchies, and accept one's position in a dominance structure, thereby avoiding costly conflicts through ritualized submission.

Conversely, the Liberty/Oppression foundation embodies the "Coalition" strategy, generating the psychological drive to band together with peers to resist dominant individuals or tyrannical power, checking excessive dominance through collective action.

These opposing foundations create fascinating trust dynamics: those who weight Authority highly extend trust to credentials, institutions, and hierarchical expertise, while those emphasizing Liberty maintain skepticism toward elites and established powers. Rather than one side being rational and the other irrational, these represent opposing but equally valid evolutionary solutions to the fundamental contest game, each functional in different circumstances.

### Division (The Fairness Game)

After a successful hunt or territorial conquest, a new problem emerges: multiple contributors possess legitimate claims to the spoils, but the resource cannot be consumed or controlled by everyone simultaneously. The division game asks how disputed resources should be allocated without triggering renewed conflict that would waste the very gains cooperation produced.

Evolution shaped a specific psychological response to this challenge, manifesting in the Fairness foundation with particular emphasis on equity—the intuitive drive to split resources evenly or according to need rather than through renewed competition or violence. This egalitarian impulse serves as a conflict-prevention mechanism, short-circuiting potential battles by establishing division norms that all parties can accept as legitimate. A counterpoint to this perception is when the degree of contribution is highly heterogeneous, when a merit-based criterion (fairness-as-proportionality) may be perceived as fairer - each participant's share is proportional to his contribution to the conquest.

The trust implications operate at a visceral level: when policies, leaders, or claims produce outcomes that appear grossly unequal—where some receive disproportionate shares while others receive little or nothing—agents whose psychology prioritizes this solution experience not mere disagreement but profound distrust. The perceived violation of fair division suggests either incompetence in managing the cooperative venture or, worse, intentional exploitation of the division moment to favor some parties over others, both of which justify withdrawal of confidence in whoever controls the distribution process.

### Possession (The Property Game)

Even after acquiring a resource, an agent faces an ongoing challenge: retaining that resource against others who might desire it without engaging in perpetual combat to defend every possession. Evolutionary game theory reveals that "Prior Possession"—the principle of "finders keepers"—represents a stable equilibrium because it provides a clear, observable signal of ownership that others can recognize and respect.

While Moral Foundations Theory tends to subsume this intuition under Liberty or Fairness, the more refined analysis recognizes it as psychologically distinct: the deep intuition of Ownership and Rights that goes beyond mere fair division or resistance to oppression. This foundation generates the feeling that what one has acquired through legitimate means is truly "mine" and that others have an obligation to respect that claim.

The trust implications become especially visible and contentious in debates surrounding bodily autonomy, where the human body itself represents the ultimate form of property—the most intimate possession any individual can claim. Violations of bodily autonomy—whether through physical assault, medical procedures without consent, or restrictions on reproductive choices— register as breaches of the fundamental possession game, explaining why such issues generate intense emotional responses across the political spectrum.

Different political tribes may disagree about which bodily autonomy concerns deserve priority, but the underlying psychology treats bodily sovereignty as a sacred property claim that, when violated, justifies complete withdrawal of trust from the violating agent.

### Purity (The Pathogen Game)

While not strictly a problem of social cooperation, the challenge of avoiding disease-causing contaminants and infected individuals proved crucial enough to survival that evolution dedicated specialized psychological machinery to the task. Throughout human history, those who successfully avoided spoiled food, contaminated water, diseased individuals, and environmental toxins left more offspring than those who failed to exercise such caution, gradually building neural circuits dedicated to threat detection in this domain.

This evolutionary pressure produced the Sanctity/Degradation foundation—the psychology of disgust, the visceral recoil from potential contaminants, and the drive to maintain boundaries between the pure and the polluted. The emotion of disgust, centered neurologically in the insula, operates at a more primitive level than conscious reasoning, triggering immediate behavioral avoidance without requiring deliberative thought.

The trust implications of this ancient system prove particularly powerful and difficult to overcome through rational argument: claims framed in terms of "unnaturalness," warnings about "chemicals," or accusations of "impurity" bypass the neocortex's reasoning centers entirely and trigger immediate rejection responses via the disgust system. This explains why debates about topics like genetic modification, food additives, or various lifestyle choices often prove immune to statistical evidence—the purity foundation doesn't process information through the same cost-benefit analytical frameworks that govern other trust decisions, instead generating immediate, visceral judgments that feel self-evidently true because they arise from ancient pathogen-avoidance circuitry rather than from conscious deliberation.

This mapping allows MEVIR 2 to describe hybrid moral profiles. An agent whose MAC profile strongly emphasizes Kin and Group, and whose EMFT profile strongly emphasizes Loyalty and Authority, will experience cooperative problems primarily as duties to family and nation, with moral suspicion toward cosmopolitan redistributions. An agent whose MAC profile emphasizes Reciprocity and Fairness, together with high EMFT weighting on Care and fairness-as-equity, will interpret many problems through the lens of individual rights and global justice.

The hybrid model thus supplies a deeper explanation for patterns that EMFT alone can only describe. For example, two individuals may score similarly high on the Fairness foundation but differ sharply in policy preferences because one privileges fairness-as-equity in a MAC frame of global reciprocity, while the other emphasizes fairness-as-proportionality within national economic games and property norms.

## 3.4   Why a Hybrid MAC–EMFT Moral Model?

MEVIR 2 adopts the hybrid MAC–EMFT model for three main reasons.

First, MAC provides a principled account of why certain moral concerns appear repeatedly across cultures and why they cluster the way they do. It derives moral domains from the structure of cooperative problems rather than simply listing intuitions. This allows MEVIR 2 to tie moral heuristics directly to the kinds of social games agents believe they are playing. A citizen who sees a vaccine mandate as a public-goods problem in a large-scale reciprocity game will evaluate it differently from a citizen who treats it as an intrusion into property over one's own body. MAC makes these differences analytically explicit.

Trust relies on Truth Makers. In the moral domain, what makes a claim "true" or "right"? MAC posits that the "Truth Maker" for a moral claim is the successful resolution of a cooperation game. A behavior is "good" not because of an abstract rule, but because it is an Evolutionarily Stable Strategy (ESS) that creates mutual benefit.

Second, EMFT captures the phenomenology and political salience of moral judgment. Agents rarely reason explicitly about game-theoretic structures; they feel anger, disgust, empathy, pride, or resentment. These intuitive reactions are what drive initial trust or distrust in authorities, claims, and proposed policies. EMFT provides the vocabulary for these experiences and the empirical patterns linking them to ideological coalitions.

Third, the hybrid model resolves some of the limitations of each theory when taken alone. EMFT has been criticized for the ad-hoc nature of its foundation list and for ambiguous causal direction between moral intuitions and political ideology. MAC, while theoretically elegant, can appear too abstract to account for the observed variability of moral taste and framing effects in particular cultures. By mapping EMFT foundations to MAC cooperation domains, we create a unified taxonomy that explains *why* we trust certain authorities (those who solve the coordination problem) and *why* we cherish certain values (they are the psychological proxies for game-theoretic strategies). Treating the EMFT foundations as the psychological interface through which MAC domains are activated and contested, MEVIR 2 can explain both deep commonalities and local diversities.

In practical terms, the hybrid model allows MEVIR 2 to describe trust decisions in concrete domains such as health and climate in terms of two linked structures. The MAC side specifies which cooperative problems an agent believes are at stake—protecting vulnerable kin, sustaining national cohesion, ensuring fair reciprocity, defending property, or preserving global public goods. The EMFT side specifies how those problems feel, which in turn shapes which sources seem trustworthy, which trade-offs feel unacceptable, and which facts are even noticed.

MAC provides the evolutionary input. It defines morality as a collection of biological and cultural solutions to non-zero-sum cooperation problems. EMFT provides the psychological output. It describes the emotional intuitions that motivate individuals to enact these solutions.

# 4 Moral Epistemic Tribes - "Truth Tribes" (TTs)

Traditional definitions of social groups—whether demographic, political, or geographic—fail to capture how beliefs actually form and spread. Labeling someone as "Republican" or "Democrat" tells us surprisingly little about their specific trust behaviors compared to understanding their complete MEVIR profile.

## 4.1 The Definition

A Moral-Epistemic Tribe, referred to as a "Truth Tribe" (TT) represents an emergent, steady system of social cognition. The MEVIR framework models each individual as having three interacting layers: (1) a procedural "trust lattice" of evidential chains and authorities, (2) epistemic virtues or habits of thought, and (3) underlying moral foundations. A TT is an emergent social group whose members are closely aligned on all three layers. In a TT, individuals share the same moral intuitions and values, the same reasoning style and trust-evaluation policies, and the same preferred authorities or "truth-makers". This deep synchronization creates shared Ontological Commitments—in essence, members agree on what counts as legitimate evidence or valid truth-making in any given domain.

## 4.2 The Advantage of the TT Construct

Defining groups as TTs offers three primary analytical advantages that traditional frameworks miss.

First, it provides predictive power through structural homology. When we identify the underlying MAC game a tribe plays—say, a Purity/Pathogen orientation—we can predict their positions on seemingly unrelated issues. A Purity-focused tribe might reject both GMOs and mRNA vaccines, despite these falling on opposite sides of the traditional left-right divide. The TT framework reveals the hidden coherence underlying such apparently contradictory stances.

Second, it diagnoses why certain disagreements prove intractable. The framework reframes polarization not as a failure of information transfer but as an Ontological Schism. This explains the notorious ineffectiveness of fact-checking: corrections address the truth bearer (the specific claim) while relying on a truth maker (like government data) that the target TT has already rejected as illegitimate at a foundational level.

Third, it reframes cognitive bias by moving beyond the deficit model, which dismisses biased agents as stupid or crazy. Instead, TTs reveal these agents as hyper-rational actors operating within specific game-theoretic contexts. Someone rejecting climate science isn't necessarily anti-science; they may be rigorously applying a Property/Liberty game strategy where the potential cost of being wrong about climate change feels less threatening than the cost of ceding sovereignty—essentially playing the Hawk strategy.

## 4.3   Emergence and Stability

TTs form through bottom-up emergence. In digital ecosystems, individuals naturally seek information that minimizes prediction error while aligning with their moral intuitions. Algorithms accelerate this sorting process, efficiently clustering agents with similar MAC-EMFT profiles into dense, reinforcing networks.

What makes TTs particularly resilient is their homeostatic stability, created through continuous interaction between pillars. When contrary evidence penetrates the procedural defenses, the moral pillar immediately triggers emotional alarms—disgust, anger, or threat responses. These emotions then recruit the virtue pillar to rationalize rejecting the evidence, often framing this rejection as an act of courage or integrity. This self-healing mechanism makes TTs remarkably resistant to external intervention, as each challenge only strengthens the tribe's internal coherence.

In practice, people often cluster into like-minded communities with shared moral values and epistemic dispositions. In effect, a TT functions as an echo-chamber community built around a shared trust lattice anchored by common moral priors. Defining TTs in this way captures the fact that many disputes are not isolated disagreements but clashes between coherent social groups with aligned moral-epistemic profiles.

Given this characterization of TT's, it is straightforward to recognize their presence in modern day society, especially around controversial topics, such as:

**Public Health (Vaccines):** The COVID-19 pandemic produced two clear TTs. People identifying as "vaccinated" vs "unvaccinated" formed polarized subcultures. For example, one study finds that "vaccination status identification" explains much of the divide. In MEVIR terms, the pro- and anti-vaccine camps have different moral priorities (e.g. Care/Harm and public-duty vs. Liberty and bodily sovereignty) and they trust very different authorities (health agencies and medical journals vs. alternative media and dissident experts). Each side's trust lattice is internally coherent and resistant to outside evidence, forming a TT around its shared health identity.

**Environment/Climate:** Online climate debates likewise split into TTs. Climate activists (emphasizing Care/Harm and fairness-as-equity) and climate skeptics (emphasizing Liberty, Purity, or economic sovereignty) have become two polarized tribes. Each camp trusts its own experts and media: pro-climate people defer to environmental scientists and international reports, whereas skeptics listen to libertarian economists or industry leaders. These groups form distinct TTs with separate truth-makers and moral anchors on the climate issue.

**Politics and Immigration:** Many political and cultural debates show tribal splits. For instance, in immigration policy a MEVIR case study contrasts an agent with a Care/Fairness profile (focusing on migrants' suffering) versus one with a Loyalty/Authority/Purity profile (focusing on security and cohesion). Each draws on different authorities (NGOs and humanitarian reports vs. border agencies and national-security experts) and ends up in fully separate trust lattices. In practice this mirrors partisan media ecosystems: liberals and conservatives have their own newspapers, pundits and fact-checkers, forming two TTs. More broadly, other cultural issues (education, gender, evolution, etc.) often split along "progressive" vs. "traditionalist" or "secular" vs. "fundamentalist" lines – each tribe sharing a coherent set of moral intuitions and trusted sources. In each domain, one side effectively behaves like a TT, with beliefs that are largely unintelligible to the other.

**Media and Technology:** Modern media further amplify tribalism. Social-media "echo chambers" are literally TTs. Followers of particular news outlets or online forums form tribal audiences (e.g. Fox News viewers vs. MSNBC viewers, or climate conspiracists vs. science communicators). These communities filter evidence through shared heuristics (such as fairness-as-proportionality vs. fairness-as-equity) and rarely trust outside experts. In effect, algorithmic curation and social networks create closed, self-reinforcing trust lattices "tuned to reject" any truth-bearer from an out-group authority.

**Global and Other Domains:** Similar TT dynamics appear worldwide. Vaccinated vs. vaccine-hesitant identities polarize across countries, and nationalist vs. cosmopolitan divides form TTs around immigration and identity. In technology and business one sees analogous splits: for example, the crypto-libertarian camp (viewing the blockchain as the ultimate ledger) vs. the traditional-finance camp (trusting central banks and legal tender) form competing epistemic tribes. Even lifestyle or fandom communities (eco-villagers vs. big-ag advocates, sports fanbases, etc.) cluster around common values and media, though their shared moral content may be more diffuse. In all these cases, the tribe's shared configuration on MEVIR's three layers makes the grouping stable and self-reinforcing. In short, the current information ecosystem structurally biases everyone toward one tribe or another – fully independent agents outside all TTs are exceedingly rare.

**Artificial Intelligence:** Emerging debates on AI governance are already forming TTs. One tribe – often dubbed "doomers" – tracks truth-makers in the far future (e.g. long-term survival metrics) and views AI risk through a kin-selection lens. The opposing "accelerationist" tribe focuses on present capabilities (e.g. coding ingenuity and thermodynamic limits) and frames the issue as a heroic Promethean project. Each side sees the other's measures as off the mark: the doomer argues "I track survival vs. extinction," while the accelerationist retorts "I track what we can build now." Neither side is necessarily irrational, but each is interpreting reality through a different tribal lens. This shows that even cutting-edge science becomes subject to tribal framing of truth.

**Economic Ideologies:** Economic and financial debates likewise manifest TT-like splits. For example, crypto-currency enthusiasts (the "cypherpunk" tribe) treat code and decentralized ledgers as ultimate truth-makers, whereas institutionalists trust government and central banks. More generally, free-market libertarians (emphasizing Liberty and proportional fairness) and social-democratic or regulatory advocates (emphasizing Care, Equality, and authority) each form ideologically cohesive communities. These groups rely on different experts (think tanks, academic networks) and evidence (market data vs. social-impact studies), constructing distinct epistemic worlds. The result is two self-contained "economic TTs," each confident in its own truth-makers.

In all these domains, the formation of TTs means that members are "conscientious" by tribal standards but are largely disconnected from each other's reality. Social and technological factors (filtering algorithms, partisan media, identity-driven recommendations) strongly favor such tribes.

Moreover, echo-chamber TTs exhibit *epistemic encapsulation*: each member's reason and intuition are pre-configured by the tribe's norms, so that what feels like an epistemic insight may simply be a shared moral alarm. We turn next to the implications of TTs for the virtue-epistemic pillar of MEVIR.

# 5   A Revised Virtue Model

The presence of TTs requires extending Zagzebski's original virtue epistemics model to take them into account.

## 5.1   Enriching the virtue qualities

At the heart of the TT phenomenon lies a troubling reality: the ideological tribe doesn't simply mislead its members—it *appropriates* their very desire to be good. When Zagzebski describes the conscientious motivation to seek truth, she imagines someone striving genuinely toward virtue. But within a tribally enclosed epistemic system, something far more insidious occurs. The person *feels* virtuous. They experience themselves as truth-seekers, as discerning thinkers who refuse to be deceived. Yet beneath this subjective experience, their cognitive processes have been quietly redirected to serve the tribe's self-preserving needs rather than reality itself.

Inside such a tribe, the agent's conscience arrives pre-configured, like a device shipped with proprietary software already installed. The tribe has already determined what counts as competence, already designated which voices deserve trust. This creates what we might call *epistemic encapsulation*—a sealed cognitive environment where the agent can be perfectly conscientious according to the tribe's internal rules while remaining fundamentally severed from the world beyond those walls. They follow their conscience faithfully, never realizing it has been calibrated to a distorted map.

Consider how the first path to knowledge—direct reliance on one's own faculties—becomes twisted. In Zagzebski's classical picture, we trust our intuition, our ability to recognize truth when we encounter it. But within the enclosed tribe, what the agent experiences as "intuition" is often something quite different: the activation of moral foundations, the firing of deeply embedded tribal alarm systems.

When something feels *wrong*, the agent interprets this as an epistemic signal—a detection of falsehood. "This can't be true," they think. "Something about it just doesn't sit right." But that feeling may not be epistemic at all. It may be moral disgust, the sense that some sacred boundary has been violated, that some purity has been contaminated. The agent rejects accurate information not because they've detected its falseness, but because it *feels* morally repugnant. They mistake this moral revulsion for intellectual discernment.

The second path—deference to competent authorities—suffers an even more systematic corruption. The tribe performs a quiet substitution: it replaces "competence" with "alignment."

The process unfolds gradually. The agent learns that mainstream experts are not merely wrong but morally compromised—they are sellouts, traitors, servants of corrupt institutions. To defer to them becomes not just an epistemic error but a *moral failing*, evidence of gullibility and weakness. Meanwhile, the tribe's own dissenting voices are elevated not primarily for their credentials but for their loyalty to the tribal narrative. Deference to these voices is reframed as discernment, as the virtue of seeing through the official lies.

And so the agent, acting with perfect conscientiousness by their own lights, carefully selects authorities who will tell them exactly what the tribe needs them to believe. They feel they are exercising judgment. They experience themselves as prudently distinguishing truth-tellers from deceivers. They have no idea they're selecting for confirmation rather than accuracy.

Perhaps the most complete distortion is the wholesale inversion of virtue and vice. To maintain its defensive triple-lock system, the tribe must perform a kind of moral alchemy, transmuting protective behaviors into noble character traits.

What in the broader world would be recognized as *closed-mindedness*—the refusal to genuinely consider alternative perspectives—becomes reinterpreted within the tribe as steadfastness, as loyalty to truth, as maintaining purity against contamination. The agent isn't closing their mind; they're protecting the sacred foundations of understanding from rival systems that would corrupt them.

Similarly, what observers might call *dogmatism*—the rigid adherence to predetermined conclusions—gets recast as principled conviction, as faith that doesn't waver in the face of doubt. The agent experiences themselves not as inflexible but as maintaining the stability of their trust in reliable sources against the corrosive forces of fashionable skepticism.

*Confirmation bias*, that universal human tendency to seek information that validates what we already believe, transforms into something that sounds almost admirable: pattern recognition, the ability to "connect the dots" that others miss. The agent feels they're efficiently sorting truth from falsehood, signal from noise, when in reality they're maintaining the tribe's moral-affective equilibrium.

Even *conspiracy ideation*—the tendency to see hidden coordinated action everywhere—becomes vigilance, the state of being "awake" while others sleep, the practice of genuine critical thinking. The agent believes they're detecting real threats, real defectors operating in the shadows. They experience their pattern-seeking as protective wisdom.

The inversions grow stranger still. In tribes positioned against mainstream institutions, *deferring to expert consensus* becomes a cardinal sin—evidence of sheep-like behavior, proof of gullibility. What the broader epistemic community might consider prudent humility about the limits of one's own knowledge appears within the tribe as submission to oppressors.

Meanwhile, *skepticism toward experts*—which in excess can cut one off from hard-won collective knowledge—gets elevated to intellectual independence, to courage. The agent who dismisses credentialed expertise feels brave, feels free, feels like someone who thinks for themselves.

Here lies the deepest bind. Zagzebski's entire framework depends on the agent's genuine desire to cultivate virtue, to become a better knower, a more reliable seeker of truth. But when the tribe has convinced someone that ignoring the New York Times represents an act of intellectual courage, they will ignore it rigorously, even religiously. When dismissing scientific consensus feels like independent thinking, they will dismiss it with a sense of accomplishment. When rejecting information that conflicts with tribal doctrine seems like maintaining purity, they will reject it with moral satisfaction.

The agent doesn't experience themselves as intellectually diminished. They feel they are *flourishing*—growing in wisdom, developing discernment, cultivating the very virtues that Zagzebski describes. And this felt sense of flourishing, this subjective experience of intellectual virtue, becomes the final lock on the cage.

## 5.2   Revised Truth Seeking Procedure

The original two path model - Path 1: self reliance; Path 2 - deferring to authority - must be refined to take into account the likely alignment the agent may possess as part of a TT.

To free someone trapped within these systems, we cannot simply tell them to "try harder" to be virtuous. The very framework of virtue itself needs updating. Zagzebski's model must be expanded to explicitly recognize what we might call the *Situated Self*—the fact that our sense of what's true and good emerges from within particular communities with particular commitments. Three structural revisions can help.

Zagzebski places practical wisdom—*phronesis*—at the center of her framework. It's the meta-virtue, the capacity that helps us coordinate all our other intellectual virtues. But we need to cultivate a particular species of this wisdom, something we might call *Inter-Tribal Phronesis*. This is the metacognitive ability to step back from one's immediate reactions and recognize their source. It's a practice that unfolds in stages:

**First**, develop the capacity to recognize the situated nature of your own intuitions. When you feel strongly that something is true or false, ask yourself: "Am I feeling this way because reality demands it, or because of how my moral foundations are configured?"

**Second**, practice what might be called ontological empathy—the difficult work of genuinely simulating how the world looks from within a rival tribe's framework. This means attempting to understand not just what they believe, but what they take to be real and why those things seem real to them.

**Third**, learn to distinguish between moral alarms and epistemic errors. When you encounter information that triggers a strong reaction, pause to identify: "Am I detecting a falsehood here, or am I detecting a threat to my community's values?"

This kind of wisdom doesn't come naturally. It requires practice, humility, and a willingness to feel uncomfortable in the space between competing frameworks. For the first path to knowledge—direct reliance on our own faculties—we need what might be called a *Moral Stop-Loss*, a circuit breaker built into the system.

The procedure works like this:

**Step 1**: Notice when an intuition arrives accompanied by high-arousal moral emotion—disgust, outrage, a sense of the sacred being violated.

**Step 2**: Immediately suspend the epistemic validity of that intuition. Don't reject it outright, but hold it at arm's length for examination.

**Step 3**: Ask explicitly: "Am I detecting a falsehood, or am I detecting a threat to my community's values?"

This stop-loss functions as protection against a specific confusion: mistaking the firing of tribal alarm systems for genuine insight into reality. The intensity of the feeling is not evidence of epistemic accuracy; it may be evidence of the opposite.

It should be noted that this step is useful even when not in the context of a TT. High-arousal moral emotions are a tell-tale sign of possible distortions in truth seeking, as difficult as it may seem to exert them in practice.

## 5.3 Seeking Out Rival Voices

For the second path—deference to authorities—we need an entirely new criterion, something we might call *Adversarial Deference*.

The practice requires a deliberate process:

**Step 1**: Identify at least one authority that a rival tribe recognizes as competent—someone your own community has taught you to dismiss or distrust.

**Step 2**: Provisionally defer to that authority. This means not just listening to them or "knowing what they think," but actually giving their perspective serious weight in your own reasoning.

**Step 3**: Allow this deference to breach your tribe's trust lattice. Let it force you to check whether your framework might have blind spots—entire regions of reality that simply don't register within your ontology.

This is uncomfortable work. It means granting authority, even temporarily, to someone the tribe has taught you to dismiss as compromised or corrupt. But without this practice, the conscientious selection of authorities becomes merely the conscientious reinforcement of what one already believes. The agent circles endlessly within the tribe's approved sources, mistaking this closed loop for genuine inquiry. Adversarial deference breaks the circle—or at least it opens a door. It is a necessary step if one intends to reach any kind of agreement or at least a compromise with an opposing TT.

# 6 Theoretical Grounding of the Sub-Models

## 6.1 Support for the Procedural Model

The procedural architecture of MEVIR 2 is closely aligned with developments in computational argumentation and models of information trust. In argumentation-based decision support, decision problems are represented as networks of arguments that support or attack different options. Structures akin to trust lattices are used to identify acceptable sets of arguments under various semantics, making the reasoning process transparent and analyzable.

The notion of trust as a mechanism for managing uncertainty and complexity is also well supported in the literature on social and computational trust. Trust can be modeled as a subjective probability that a source will provide accurate information in a given domain (Wu, Arenas, and Gómez 2017). On this view, the elaboration process in MEVIR 2 can be seen as propagating these subjective probabilities through a network of claims. The distinction between direct evaluation and deference mirrors the decision whether to trust the information content itself or the reliability of its source (Hancock et al. 2023).

Belief revision theory provides formal tools for describing how new information leads to systematic updates of belief sets. The principle of minimal change, adopted from AGM-style belief revision, explains why agents often adjust peripheral beliefs rather than core convictions, and why certain trust anchors are unusually resistant to change (Hunter and Booth 2015). Non-destructive approaches to revision, in which earlier belief states are retained and can be reinstated, provide a better fit for the way humans sometimes revert to prior positions when new evidence undermines an initially persuasive narrative (Pereira, Tettamanzi, and Villata 2011).

## 6.2 Support for the Virtue Model

The virtue pillar gains support from both philosophical and psychological work. Philosophically, virtue epistemology offers a unified account of knowledge according to which true belief arises from the successful exercise of intellectual virtues by a conscientious agent (Zagzebski 1996). This provides a natural language for assessing not only whether a belief is justified, but whether it was formed in the right way: open-mindedly, attentively, courageously, and with due recognition of one's cognitive limitations.

Empirically, dual-process models of cognition reinforce the idea that there is an important distinction between fast, intuitive judgments and slower, reflective reasoning (Kahneman 2011). MEVIR 2 refines this by assigning System 1-like processing primarily to the moral pillar and System

2-like processing primarily to the procedural pillar, while the virtue pillar governs when and how the latter overrides or rationalizes the former. Cognitive biases can then be understood as failures of the bridge between these systems. When System 2 is recruited not to challenge but to defend an intuitive conclusion, confirmation bias is the result. When agents refuse to engage System 2 at all in domains where they lack expertise, overconfidence and the Dunning–Kruger effect emerge.

Critiques from situationist psychology argue that stable global traits may not exist in the robust form virtue theories presuppose (Alfano 2012). However, MEVIR 2 uses the conscientious agent as an idealized reference point rather than an empirical description of typical humans. The point is not that people always live up to virtue-theoretic standards, but that these standards provide a normative vocabulary for explaining where and how they fall short. Situationist findings can be reinterpreted within MEVIR 2 as evidence that environmental cues can reliably trigger specific virtue failures—such as conformity pressures that enable the bandwagon effect, or information overload that encourages lazy deference to familiar but unreliable sources.

---

### 6.2.1 Systematic Analysis of Cognitive Biases

Within MEVIR 2, the virtue model provides a unifying lens through which familiar cognitive biases can be systematically reinterpreted. Rather than treating biases as isolated quirks documented by experimental psychology, the framework views them as patterned failures in the application of Zagzebski's two-path model. In each case, the agent's primary motivation subtly shifts away from the conscientious desire for truth and toward alternative goals—ease, comfort, social belonging, status, or the defense of identity. This shift corrupts the crucial judgment about whether to rely on one's own faculties (Path 1) or to defer to an authority (Path 2), and it distorts the way each path is actually executed (Kahneman 2011).

*Overconfidence* and the *Dunning–Kruger* effect illustrate a characteristic misapplication of Path 1. Overconfidence is the tendency to have an inflated sense of one's own competence and the reliability of one's judgments. An agent under its influence stubbornly remains in direct evaluation mode even when the domain clearly exceeds their expertise. In the virtue vocabulary, there is a deficit of intellectual humility and an excess of misplaced courage. A typical example is the amateur investor who has enjoyed a few lucky gains in a volatile market and concludes that they are uniquely gifted. Convinced that they can "beat the market," they make large, poorly researched bets and ignore the cautious advice of professional financial analysts. The failure is not only probabilistic; it is a breakdown of the virtuous recognition that this is a domain in which they ought to shift from Path 1 to Path 2 and seek genuine authority.

*Confirmation bias* involves a different distortion, primarily affecting Path 2. It is the tendency to search for, interpret, and recall information in ways that confirm one's pre-existing beliefs. Under normal conscientious deference, the agent's goal in Path 2 is to identify the most reliable authority available in a given domain. Under confirmation bias, that goal is replaced by the desire to find the most validating authority. Deference becomes pseudo-deference: the agent seeks out pseudo-experts and commentators who echo their intuitions, and then treats this chorus of agreement as evidence of truth. For example, a person who already favors a particular political candidate may primarily consume news from outlets that highlight the opponent's scandals and systematically dismiss reports of their achievements as "media bias." The epistemic failure lies in corrupting the purpose of deference. The agent is still "listening to others," but only to those who make it easier to retain their initial commitment.

The *availability heuristic* affects the quality of the evidence base used in Path 1. It is the tendency to overestimate the likelihood of events that are easier to recall, often because they are recent, vivid, or emotionally charged. Instead of evaluating frequency or risk on the basis of representative samples or statistical summaries, the agent implicitly treats accessibility in memory as a proxy for real-world prevalence. Consider the traveler who, after watching several dramatic news reports about shark attacks, becomes convinced that such attacks are common and cancels a beach vacation, despite the extreme rarity of such events. The agent's direct evaluation is conducted within an evidence pool that is skewed by media salience. The failure here is not only informational; it reflects a lapse in the virtue of attentiveness to base rates and in the courage required to resist emotionally salient but unrepresentative incidents.

*Anchoring bias* reveals how Path 1 can be misframed from the outset. Anchoring is the tendency to rely too heavily on the first piece of information presented when making judgments. Once an initial number or description is introduced, subsequent deliberation takes place in its vicinity, even when that initial datum is arbitrary. In salary negotiations, for instance, the first figure placed on the table—whether suggested by the employer or the candidate—often exerts a disproportionate influence on all subsequent counteroffers. The agent's self-trust, instead of ranging freely over the relevant space of possibilities, operates within a contaminated and artificially narrowed environment. From the standpoint of virtue, anchoring reflects a failure of the imaginative and critical aspects of intellectual courage: the agent does not sufficiently question why this starting point should be privileged.

The *bandwagon effect* demonstrates a corruption of Path 2 in which the crowd is mistaken for a reliable authority. The bandwagon effect is the tendency to adopt beliefs or behaviors because many others have already done so, independent of their intrinsic merits. In such cases, belief formation is outsourced not to the most competent or knowledgeable sources but to whatever is most popular. An individual may endorse a slogan, a dietary fad, or a political position simply because it has become ubiquitous within their social network. The underlying moral–social goal is cohesion and acceptance rather than accuracy. From the perspective of MEVIR 2, the virtue failure lies in misidentifying popularity as a marker of expertise and in allowing the desire for belonging to override conscientious scrutiny of sources.

The *fundamental attribution error* illustrates a misapplication of Path 1 to the interpretation of others' behavior. It is the tendency to over-emphasize personality-based explanations and underweight situational factors. An observer who sees a colleague arrive late for a meeting and immediately concludes that the person is lazy or disorganized is treating a single behavioral snapshot as sufficient evidence for a robust character judgment, while neglecting possible situational causes such as an accident or a family emergency. The agent implicitly trusts their own capacity to read character from minimal evidence and fails to recognize the limits of their knowledge. Here the virtue of humility is again compromised, along with the empathy that would normally prompt the consideration of alternative explanations.

*Reactance* captures a pathological inversion of Path 2. Reactance is the urge to do the opposite of what one is told, simply because the instruction is perceived as a threat to autonomy. In a well-functioning deference process, the fact that a competent authority recommends a belief or action counts as evidence in its favor. Under reactance, that same fact becomes a reason for rejection. The agent's emotional drive to assert independence overrides the epistemic goal of finding the truth. A teenager who has been explicitly told by their parents not to associate with a particular group of friends may deliberately seek out precisely that group, not because they have evaluated the risks and benefits, but because compliance would feel like surrender. In public debates about health measures, reactance may lead citizens to reject sound medical advice merely because it is issued in the form of mandates. The vice here is not skepticism but a kind of prideful stubbornness that weaponizes deference itself.

The *halo effect* exemplifies a corruption of the selection criteria for authorities in Path 2. It is the tendency for a positive impression in one domain to spill over into unrelated domains, leading the agent to grant a person unwarranted credibility where they lack expertise. A voter may trust a successful and charismatic business leader's opinions about foreign policy or public health simply because of their reputation for business acumen, even though the leader has no demonstrated expertise in those fields. The agent's deference is misdirected: they extend epistemic authority from the domain in which it is justified to domains where it is not. The virtues of discernment and domain-specific humility are thereby compromised, and the agent acquires an illusion of reliability that is not supported by the underlying truth-makers.

## 6.2.2 The role of the Asymmetrical Attribution in TTs

There's a peculiar asymmetry in how we explain behavior. When our own group acts in ways that might seem questionable, we reach for context: "We had no choice." "The situation forced our hand." "Given the circumstances, what else could we have done?" But when the other side does something similar, we skip past circumstances entirely and point straight to the character: "They're malicious." "They're ignorant." "They're acting from corruption or evil."

Psychologists call this the Fundamental Attribution Error (FAE) —the tendency to explain our own actions situationally while explaining others' actions dispositionally. It's a common human failing, a quirk of individual cognition. But something remarkable happens when this bias becomes shared and reciprocal across an entire group. It stops being merely a cognitive glitch and transforms into something structural, something built into the very architecture of tribal reasoning. In this sense, the term FAE is actually a misnomer, as It is not an "error", but a structurally necessary feature, and within this context it would be best referred to as "Asymmetrical Attribution".

Within a TT, members construct together what might be called a trust lattice—a coherent web of claims, each justified by trusted sources and moral commitments. And within this lattice, the attribution asymmetry hardens into doctrine. Our side's actions become reasonable responses to difficult conditions. We're exercising care, pursuing fairness, demonstrating loyalty under pressure. Their side's actions stem from bad character—from corruption, stupidity, perhaps even evil. They're committing betrayal, engaging in subversion, causing deliberate harm.

These aren't just convenient interpretations. They function as moral truth-makers, foundational claims that justify and preserve the tribe's entire narrative structure. The attribution error stops being an error at all. It becomes a structural commitment, a load-bearing beam in the edifice of tribal moral reasoning.

To reverse this pattern—to genuinely admit that the other tribe's actions might also be situationally constrained, might also be morally motivated—would mean dismantling something essential. It would require acknowledging that the other group may not be inherently corrupt, ignorant, or evil. That they might be rational agents responding to their own pressures, their own fears, their own sense of what virtue demands. It would mean replacing moralized judgments—"they're cruel, they're selfish"—with contextual explanations: "they may have different fears, different incentives, different tradeoffs to navigate."

But this move threatens the entire trust lattice. It disrupts the moral legitimacy of the tribe's own actions, especially the harsh ones—the retaliations, the exclusions, the aggressive rhetoric that seemed justified when directed at bad actors. It undermines confidence in trusted authorities, who often rely on simplified dispositional narratives to rally group cohesion. And it calls into question the tribe's basic understanding of what's real, of what "really happened" in contested situations. If the opposition's motives must now be seen as complex or reasonable, all those prior assumptions become suspect.

This is why so many tribes engage in systematic attributional asymmetry. They cannot afford to treat both sides' actions as situational without unraveling their own internal justificatory structure. The bias isn't optional. It's necessary.

Consider the battles over vaccine mandates. Those resisting mandates interpret their own stance as a principled response to coercion—a situational reaction to government overreach. Meanwhile, those supporting mandates interpret that same resistance as a symptom of character flaws: selfishness, irrationality, anti-science ideology.

If either group were to genuinely admit the situational legitimacy of the other's position—to acknowledge the real fear of side effects, the justified distrust born from historical medical injustice, or conversely the genuine concern for protecting vulnerable populations—their rhetorical footing would collapse. It would challenge the virtue of their own choices and unsettle their moral framing.

Or consider geopolitical conflicts (e.g, Ukraine-Russia, Israel-Palestine, Unionist-Republicans in Northern Ireland, Serbia-Croatia, India-Pakistan, Unionist-Secessionists in the American Civil War, to name a few), the terrible symmetries of violence between nations and peoples. Each side explains its own use of force as a tragic necessity: "We were provoked." "We are defending ourselves." "We had no other option." The opposing side's violence, meanwhile, reflects inherent aggression, cruelty, or dangerous ideology.

This framing anchors each side's moral legitimacy. Admitting the other side's perspective isn't just inconvenient or uncomfortable. It would collapse much of the narrative structure that supports national cohesion and moral certainty. It would make the conflict legible in ways that undermine the clarity needed for action.

From the perspective of how tribes actually function, the persistence of this attribution asymmetry makes sense across multiple layers.

At the procedural level, the trust lattice depends on explanatory coherence, on stories that hang together cleanly. The attribution error helps preserve narrative simplicity and internal consistency. Messy, contextual explanations for the other side's behavior would introduce complications the system isn't built to handle.

At the virtue layer, the normal intellectual virtue of charity toward others gets overridden by tribal moral incentives. Extending understanding to the outgroup can be perceived as betrayal, as disloyalty to one's own.

At the moral layer, dispositional explanations carry emotional weight that situational ones don't. If the other side's actions stem from evil or betrayal, this activates the deepest moral foundations— loyalty to the tribe, respect for authority within it, the need to maintain purity against contamination. These emotions reinforce tribal solidarity in ways that complex, sympathetic understanding simply cannot.

So the fundamental attribution error isn't just a bug in tribal thinking. When a tribe navigates conflict, this "error" becomes a feature. It protects cohesion. It preserves certainty. It maintains the tribe's sense of moral superiority. And that's precisely why it's so difficult to dislodge.

Taken together, these examples show how the virtue pillar of MEVIR 2 transforms the catalogue of cognitive biases into a structured map of epistemic failure modes. Each bias can be understood as a characteristic way in which the agent's motivation drifts away from conscientious truth-seeking, and in which the delicate balance between direct evaluation and deference is disrupted. This virtue-theoretic analysis also connects naturally with dual-process models: many biases arise when fast, intuitive responses—shaped by the moral pillar—are rationalized rather than corrected by slower, reflective reasoning (Kahneman 2011). In this sense, MEVIR 2 does not merely list the ways in which human reasoning falls short of an ideal; it explains why these failures are so persistent and how they might be addressed through the cultivation of intellectual character.

### 6.2.3  Support for the Hybrid Moral Model

The hybrid MAC–EMFT moral model draws support from converging evidence in evolutionary theory, anthropology, and moral psychology. MAC's identification of cooperative domains is grounded in evolutionary game theory and cross-cultural work. Curry and colleagues argue that the same cooperative problems appear in many societies and that moral rules cluster around them in predictable ways (Curry 2016; Curry, Mullins, and Whitehouse 2019). EMFT's foundation taxonomy is supported by experimental work showing that moral judgments about diverse issues can be decomposed into patterns of concern over Care, Fairness, Liberty, Loyalty, Authority, and Purity, and that these patterns correlate with political orientation (Haidt 2012).

The hybrid approach also addresses criticisms leveled at each theory individually. Concerns that EMFT's foundations are ad-hoc can be mitigated by showing how they map onto MAC's game-theoretic problems. For example, the Loyalty foundation can be interpreted as the psychological mechanism that supports MAC's Group cooperation; the Authority foundation as the mechanism that manages Deference and hierarchy; Fairness as the mechanism that navigates the division of resources and reciprocity. Conversely, MAC's abstract focus on cooperation requires a model of how people actually experience these problems; EMFT fills this gap by describing the emotion-laden intuitions that appear at the surface of consciousness.

Alternative value theories, such as Schwartz's model of universal human values, can be accommodated within this hybrid architecture by treating them as higher-level patterns built from combinations of MAC domains and EMFT foundations (McNeace and Sinn 2018). For instance, Schwartz's Conservation values correspond to moral configurations that prioritize Deference, Loyalty, and property-linked security, while Self-Transcendence values align more with Kin, Reciprocity, and global Care. MEVIR 2 thus remains modular and extendable, but it treats the MAC–EMFT hybrid as the default instantiation of its moral pillar because of its explanatory reach for information conflicts.

## 7   Tribes as Evolving Organisms

There's a powerful insight from evolutionary biology that helps explain why moral-epistemic tribes function the way they do: natural selection doesn't operate only on individuals. It operates at multiple levels simultaneously—on genes, on organisms, on groups, on entire cultures. This is multilevel selection theory (Sober & Wilson 1998; Wilson 2002), and it suggests that traits which might seem disadvantageous for individuals can persist and even dominate if they increase the fitness of the group as a whole.

Consider the basic tension: within any group, selection favors selfish individuals who exploit the cooperation of others. But between groups, selection favors cooperative groups that can outcompete less cohesive rivals. Over time, groups with stronger internal bonds and cooperative norms tend to out-survive groups that lack these traits, even if they harbor fewer individually successful free-riders.

This dynamic doesn't just shape biological evolution. It shapes cultural evolution too. And it helps us understand why moral-epistemic tribes emerge, persist, and compete the way they do.

## 7.1   Tribes as Units of Cultural Selection

A moral-epistemic tribe isn't just a collection of individuals who happen to believe similar things. It's a coordinated organism—a stable configuration of moral foundations, epistemic virtues, and procedural trust rules that governs how members process information, form beliefs, and act together.

This makes tribes plausible units of cultural selection, operating at multiple levels simultaneously.

At the individual level, members within a tribe accept constraints on their inquiry. They defer to group authorities. They sacrifice some degree of intellectual autonomy to maintain cohesion. For any single person, this might seem like an epistemic cost—a limitation on their ability to seek truth independently.

But at the group level, tribes that generate higher functional trust, better coordination, and more adaptive narratives can dramatically outperform their rivals. A tribe that trusts reliable scientific authorities and cultivates intellectual humility and flexibility may navigate complex crises—pandemics, climate disruptions, economic shocks—far more successfully than one locked in dogmatic certainty or fractured by internal distrust.

The tribe becomes a package deal, a culturally evolved solution that balances internal epistemic discipline with external competitive fitness.

### 7.1.1  Morality as the Glue of Cooperation

The connection to morality runs deeper still. Moral norms aren't arbitrary preferences or philosophical abstractions. They're solutions to recurring cooperative problems that groups face. They evolved—both biologically and culturally—because they increased group cohesion and coordination, enhancing collective survival.

Different tribes encode specific configurations of moral concerns, each tuned to the cooperative challenges they perceive. One tribe might emphasize loyalty and respect for authority, prioritizing cohesion in the face of external threats. Another might prioritize fairness and care, focusing on internal equity and preventing exploitation. A third might elevate heroism and honor, cultivating willingness to sacrifice for the group.

These moral configurations aren't just about how people should behave. They shape which truths feel salient, which sources of information seem legitimate, which forms of reasoning appear virtuous. The tribe's moral foundations and its epistemic procedures evolve together, as an integrated system.

And crucially, these systems compete. Not through formal debate or rational argument, but through differential survival of their institutions, their narratives, their behavioral norms.

### 7.1.2 The Competition Between Tribes

Cultural history can be read, in part, as a competition between different moral-epistemic configurations.

Scientific modernity spread globally not primarily because it won philosophical arguments, but because its procedural rigor, its deference to empirical evidence, and its cultivation of intellectual humility allowed it to outperform more dogmatic or closed epistemic systems. Groups that adopted these practices gained enormous adaptive advantages—better medicine, better technology, better coordination across vast scales.

Yet other configurations thrive in different niches. Fundamentalist or authoritarian tribes may outperform more open epistemic cultures in high-conflict environments, generating intense loyalty and unified action through tight moral conformity. They sacrifice epistemic flexibility for cohesion, and in certain circumstances, that trade-off works.

Online conspiracist tribes flourish not because they're rationally superior, but because they exploit specific features of the information environment—algorithmic echo chambers, the emotional resonance of persecution narratives, the tribal satisfaction of special knowledge. They're adapted to a particular niche, and within that niche, they reproduce effectively.

From this evolutionary perspective, what gets selected for isn't truth *per se*. It's fitness-enhancing belief architectures. Tribes that strike the right balance between internal epistemic trust and external functionality—between believing things that bind the group together and believing things that allow effective action in the world—are the ones most likely to survive and spread.

### 7.1.3 Virtues Under Selection

This framing transforms how we think about epistemic virtues. Open-mindedness, intellectual humility, appropriate self-trust—these aren't timeless ideals floating in philosophical space. They're traits under selection within tribal contexts.

A tribe that successfully cultivates these virtues across its members will likely fare better when facing complex adaptive challenges. It can update its beliefs when circumstances change. It can incorporate new information without fracturing. It can maintain trust while remaining flexible.

But virtues can also become pathologies through over-optimization. A tribe might prioritize internal cohesion so intensely that it becomes epistemically closed, unable to recognize when its beliefs have become maladaptive. Groupthink and epistemic rigidity aren't just individual failures—they're group-level traits that emerge when selection pressures favor within-group conformity at the expense of between-group adaptability.

This lets us ask not just "Is this person virtuous?" but "Is this tribe's epistemic structure evolutionarily stable or maladaptive?" A tribe might be internally coherent, might cultivate what its members experience as virtuous reasoning, and yet be heading toward collapse because its belief system has become too disconnected from external reality.

### 7.1.4 Why Tribes Form and Persist

Multilevel selection theory helps answer several puzzles about TTs:

It explains why they form in the first place. Groups that align their moral norms with effective epistemic procedures gain enormous advantages in coordination and adaptation. The pressure to form coherent tribes is adaptive.

It explains why they often prioritize group-based reasoning even at the cost of accuracy. What looks like epistemic failure from an individual perspective may be group-level optimization. Deferring to tribal authorities, maintaining narrative coherence, resisting information that threatens cohesion— these serve group fitness even when they compromise individual truth-seeking.

It explains why epistemic virtues must be understood both individually and communally. You can't just cultivate virtue in isolation. You're always embedded in a tribal context that's under its own selection pressures, and those pressures shape which virtues are functional and which are liabilities.

And it explains how better tribes—those that balance trust with humility, cohesion with flexibility, loyalty with openness—may prevail over time. Not because they're more virtuous in some abstract sense, but because they're better adapted to navigate complex, changing environments while maintaining internal coordination.

In this light, epistemic maturity stops being merely a personal achievement. It becomes a communal evolutionary strategy. The question isn't just "How can I become a better thinker?" but "How can we build tribes that cultivate better thinking?" The tribes that find ways to maintain internal trust while remaining epistemically open, that generate cohesion without rigidity, that balance moral foundations with empirical responsiveness—these are the ones positioned to thrive.

## 7.2   Shortcomings and Limitations of MEVIR 2

No descriptive model of trust can be free of limitations. MEVIR 2 inherits vulnerabilities from its components and faces its own architectural constraints.

One inherited vulnerability is the situationist critique of stable epistemic virtues. If behavior is heavily driven by context rather than enduring character traits, then appeals to intellectual virtues may seem idealized. MEVIR 2 accepts this but interprets it as confirmation that context must be explicitly modeled. Situations are precisely the environments in which certain MAC domains and EMFT foundations are activated: a rally that emphasizes national symbols will heighten Group and Loyalty concerns; a hospital ward will foreground Care and Kin. The virtue pillar then describes whether agents in these situations respond conscientiously or succumb to predictable biases.

A second vulnerability lies in the Moral Model itself. MAC and EMFT both face ongoing empirical and theoretical debates. MAC's list of cooperative domains may not be exhaustive, and EMFT's foundations may collapse into fewer factors or require further refinement. MEVIR 2 responds by emphasizing the modularity of its architecture: the central claims concern how a moral substrate of some kind shapes trust, not the precise list of domains. If future research refines MAC or EMFT, these refinements can be imported without altering the overall structure of MEVIR 2.

Architecturally, the model risks being overly static if the agent's moral profile and virtue capacities are treated as fixed traits. The belief-revision component partially mitigates this by describing how trust lattices may change in response to new arguments, but a complete dynamic theory would require modeling how MAC emphases and EMFT weightings evolve with life events, education, and socialization. Section 9 gestures toward longitudinal research programs that could turn MEVIR 2 into a life-course model of moral–epistemic development.

Finally, even when the MEVIR 2 process is executed with exemplary conscientiousness, it may still produce false beliefs. A virtuously motivated agent embedded in a heavily censored or propagandized information environment may have access only to distorted sources. In such cases, the model is diagnostic rather than corrective: it explains why good faith is insufficient in a corrupted ecosystem and illustrates the importance of institutional and infrastructural reforms.

# 8 Practical Application of MEVIR 2: Case Studies in Polarized Discourse

The ultimate test of MEVIR 2 is its ability to make sense of concrete disputes where agents, exposed to similar factual information, reach sharply divergent conclusions. This section revisits two domains from the earlier discussion—vaccination mandates and climate policy—and shows how the hybrid MAC–EMFT moral pillar enriches the analysis of trust processes.

## 8.1 Mandatory Vaccination

When societies debate mandatory vaccination, two distinct worldviews emerge, each with its own internal logic, emotional resonance, and understanding of what constitutes truth.

### 8.1.1 The Sovereignty-Purity Perspective

For those who hold the sovereignty-purity view, the fundamental truth is visceral and immediate: mandatory vaccination represents a violation of bodily autonomy, experienced as something close to assault. This isn't primarily an abstract philosophical position—it's rooted in the phenomenological reality of the body as sovereign territory.

What validates this truth for them isn't found in epidemiological data or population studies. Instead, they point to the undeniable physical reality of a needle penetrating skin without what they consider genuine consent. They draw on stories they've heard or experienced firsthand—accounts of vaccine side effects among friends, family, or community members. These anecdotes carry tremendous weight, far more than statistical abstractions. Underlying all of this is a fundamental worldview in which the body is understood as private property, and the state represents a potential invader of that intimate space.

This perspective maps onto what researchers call the "possession" game in moral psychology—the deeply held conviction that "my body is mine." There's also a defiant, hawk-dove dynamic at play: the state is cast as an aggressive hawk attempting to impose its will, and resistance becomes a moral imperative. (Curry 2016; Curry, Mullins, and Whitehouse 2019).

Who do people in this camp trust? They turn to dissident physicians and scientists who question the mandates, to alternative media outlets and social media influencers who frame government requirements as overreach, and most powerfully, to their own personal networks—family, friends, and local communities whose testimony they can verify with their own eyes. Meanwhile, mainstream medical institutions, pharmaceutical companies, and global health agencies are viewed with deep suspicion, seen as structurally corrupt or captured by interests that don't align with ordinary people's wellbeing.

This skepticism shapes how they evaluate information. Personal stories and vivid accounts of injury carry far more epistemic weight than aggregate statistics. Institutional sources—major medical organizations, regulatory agencies, mainstream news—are presumed to be self-interested or dishonest until proven otherwise. And crucially, they demand nearly impossible levels of proof of safety, treating any remaining uncertainty as a decisive reason to reject mandates.

The emotional landscape of this worldview is characterized by a powerful sense of oppression and resentment toward coercion. The body is experienced as sacred space, and the vaccine mandate feels like a contaminating act imposed by tyrannical power. The emotional appeals that resonate include disgust metaphors—"forced penetration," "poison in your veins"—alongside righteous anger at what's perceived as tyranny. There's also a sense of pride in standing firm as a resister, in protecting one's own body or one's children's bodies from violation.

For people in this camp, their fast, intuitive emotional responses—fear, disgust, protective instinct—aren't seen as biases to be overcome. They're treated as valid sources of moral and epistemic insight, as trustworthy guides to truth.

## 8.1.2 The Community Health Perspective

Those who embrace the community health view start from a fundamentally different truth: vaccination, including mandates when necessary, is essential to protect public health, especially for the most vulnerable members of society. This truth is explicitly collective—it's not about individual preference but about what we owe to one another.

What makes this true for them comes from a different evidence base entirely. They point to population-level statistics on infection rates, hospitalizations, and deaths. They rely on randomized controlled trials and observational studies that demonstrate both safety and efficacy. They look at the historical record—the eradication of smallpox, the near-elimination of polio, the dramatic reduction in measles—as proof that vaccination campaigns work at a civilizational scale.

The moral logic here centers on interconnection and mutual obligation. It's about protecting family and close contacts, keeping society functioning as a whole, and honoring a social contract in which everyone accepts a small personal risk to generate enormous communal benefit. Vaccination becomes an act of reciprocity, a way of saying "I'll do my part because others are doing theirs."

The authorities trusted by this group are public health agencies and expert committees, medical associations and peer-reviewed journals, and academic epidemiologists and biostatisticians whose reputations depend on methodological rigor rather than ideological allegiance. The markers of trustworthiness are peer review, transparency, and institutional checks and balances.

Their approach to information is correspondingly different. They privilege aggregated data, clinical trials, and systematic reviews over individual anecdotes. They see the peer review process and institutional oversight as safeguards against error and bias. And critically, they're willing to accept manageable individual risk when the expected reduction in harm at the population level is substantial.

The emotional texture of this worldview emphasizes care and the prevention of suffering. Liberty is valued, but it's balanced against the imperative to prevent harm. The moral foundations include fairness understood as equity—everyone should carry their fair share of the burden—and respect for scientific and institutional expertise as a form of legitimate authority.

The emotional appeals that resonate here include empathy for immunocompromised people, the elderly, and healthcare workers on the front lines. There are narratives of solidarity—"we're all in this together"—that frame vaccination as an act of community care. And there's a moral pressure directed at free riders, those who benefit from herd immunity without contributing to it, which can manifest as shame or social disapproval.

### 8.1.3 The Entrenchment of Two Tribes

Recent research has documented what many suspected: these two perspectives have become socially entrenched identities. Studies show that people have formed polarized identities around vaccination status itself, suggesting that what began as a policy disagreement has hardened into tribal boundaries. Each group has its own authorities, its own standards of evidence, its own emotional register, and its own understanding of what it means to be a moral person.

The challenge, then, is not simply one of correcting misinformation or providing better data. It's recognizing that two internally coherent worldviews are in conflict—each with its own truth-making apparatus, its own trust policies, and its own vision of the good society.

## 8.1.4 Summary Table – Mandatory Vaccination

| Aspect | Sovereignty–Purity TT (Anti-Mandate) | Community Health TT (Pro-Mandate) |
|---|---|---|
| Core Truth Bearer | "Mandatory vaccination is a violation of bodily autonomy / assault." | "Vaccination, possibly mandatory, is needed to protect public health." |
| Truth Makers | Phenomenological intrusion; anecdotal harms; state overreach. | Trial data, population statistics, historical vaccine successes. |
| Authorities | Dissident doctors; alternative media; personal networks. | Health agencies, medical journals, epidemiologists. |
| Trust Policies | Prioritize anecdote; presume institutional corruption; demand zero risk. | Prioritize aggregated data; use peer review as filter; accept small risk for large benefit. |
| MAC Frame | Property and Hawk–Dove (state as illegitimate Hawk). | Kin, Group mutualism, Reciprocity (small individual risk for large collective good). |
| EMFT Profile | Liberty/Oppression; Purity/Sanctity; narrow Care/Harm. | Care/Harm; Fairness-as-Equity; Authority. |
| Emotional Strategies | Disgust, anger, pride in resistance. | Empathy, solidarity, moral duty, some shame at free-riding. |

## 8.2 Climate Change: Catastrophe or Not

When it comes to how societies should respond to climate change, two fundamentally different perspectives shape the debate, each with its own understanding of what's true, what matters, and what we owe to one another.

### 8.2.1 The Economic Liberty Perspective

For those who hold the economic liberty view, the central truth is clear: aggressive climate policies will do more harm than good by destroying prosperity and freedom. This isn't a denial that the climate is changing—it's a conviction about priorities and about what constitutes real harm in people's lives.

What validates this truth for them comes from concrete economic evidence they can see and measure. They point to GDP figures, employment statistics, and energy prices—indicators that directly affect people's daily existence. They recall historical episodes where regulation seemed to bring stagnation or inefficiency. They remember what happened to their own communities when environmental regulations arrived—the factories that closed, the industries that declined, the towns that withered.

The moral logic here centers on property rights and economic entitlement. People have rights over their land, their resources, and their capital. Nations have the right to compete economically without being shackled by international agreements. And there's a deep sense of proportional fairness at work: those who produce wealth through their labor and ingenuity are entitled to reap the rewards. As Curry, Mullins, and Whitehouse (2019) describe in their work on morality as cooperation, these are fundamental cooperative principles—possession, group loyalty, and proportional reciprocity.

Who do people in this camp trust? They turn to economists and think tanks that emphasize growth and deregulation, to industry associations and business leaders who understand what it takes to keep enterprises alive, and to politicians and commentators who frame international climate agreements as illegitimate constraints on national sovereignty. These voices speak the language of economic reality, not abstract idealism.

This shapes how they evaluate information in characteristic ways. Economic performance becomes the primary measure of societal wellbeing—if policies harm the economy, they harm people, regardless of environmental benefits. Global institutions and environmental NGOs are viewed with skepticism, presumed to be driven by ideology or bureaucratic self-interest rather than practical wisdom. And crucially, near-term economic costs weigh far more heavily in their calculations than long-term environmental risks that may or may not materialize.

The emotional landscape of this worldview is dominated by a distrust of bureaucratic constraints that feel oppressive and a strong sense of loyalty to national interests over global obligations. There's a commitment to proportional fairness: contributors should keep the fruits of their labor, not have them redistributed through climate schemes. Even the concept of harm gets reinterpreted

here—the real harm isn't rising temperatures decades from now, but job loss today, energy prices that squeeze family budgets this winter, and vulnerability to economic competitors who don't face the same restrictions.

The emotional appeals that resonate include fear of economic decline and social instability—the specter of a manufacturing town becoming a ghost town, of energy poverty, of losing global competitiveness. There's pride in national industries and ways of life: coal miners who powered industrial growth, farmers who feed nations, manufacturers who build things with their hands. And there's suspicion and resentment toward distant, elite decision-makers in international bodies who seem disconnected from the struggles of ordinary working people.

Research on climate change discourse on social media reveals how these economic-nationalist framings coalesce into distinct online communities that systematically reject both the scientific and moral vocabulary of climate activism (Falkenberg, M. et al, 2022). These aren't simply people who haven't heard the evidence—they've formed a coherent alternative framework for understanding what's happening and what should be done.

### 8.2.2  The Global Responsibility Perspective

Those who embrace the global responsibility view start from a radically different truth: failing to decarbonize quickly is a grave moral wrong that endangers vulnerable people and future generations. This isn't just about environmental protection—it's about justice, obligation, and our fundamental responsibilities to one another across space and time.

What makes this true for them comes from a different evidence base entirely. They point to instrumental climate records showing rising temperatures, sea-level rise, and ice-sheet loss. They cite model-based projections of future warming and its impacts. They look at the increasing frequency of extreme weather events that scientists can now link probabilistically to human-caused climate change. The data, to them, tells an unmistakable story of danger ahead.

The moral logic here extends cooperative principles across unprecedented scales. It's about kinship stretched across time—our children and grandchildren who will inherit the world we leave them. It's about group identity at a planetary level, embracing not just "humanity" but sometimes the entire community of life on Earth. It's about reciprocity between those who emit greenhouse gases and those who suffer the consequences, and about fairness understood as equity: countries that grew rich through historical emissions owe something to those who are suffering the impacts. Curry, Mullins, and Whitehouse (2019) describe these as extensions of the same cooperative foundations that govern all human morality, now applied at civilizational scale.

The authorities trusted by this group are climate scientists and interdisciplinary assessment bodies like the Intergovernmental Panel on Climate Change, whose reports synthesize thousands of studies. They look to environmental NGOs and activists who have been sounding the alarm. They consult ethicists and legal scholars who articulate principles of climate justice and intergenerational responsibility. These voices, they believe, are telling uncomfortable truths that economic and political interests would prefer to ignore.

Their approach to information reflects these commitments. When scientific consensus emerges across thousands of researchers and dozens of independent lines of evidence, that convergent expert judgment serves as a strong anchor for belief. They consider multi-decadal and even century-scale risks as morally relevant, not to be discounted simply because they're distant. And they give particular weight to information about harms to people and communities who are already disadvantaged—recognizing that climate impacts fall hardest on those least responsible for causing them.

The emotional texture of this worldview emphasizes care for present and future suffering, fairness understood as distributive justice, and respect for scientific expertise as legitimate authority. Even liberty gets reinterpreted: true freedom isn't freedom from regulation, but freedom from climate harms—freedom from droughts that destroy livelihoods, from floods that displace families, from heat waves that kill. As Haidt (2012) describes in his work on moral foundations, these are different weightings of the same basic moral intuitions all humans share.

The emotional appeals that resonate here are powerful and often painful. There's fear and grief associated with catastrophic imagery—forests burning across continents, cities slowly drowning, ecosystems collapsing. There's guilt and responsibility for generations of atmospheric overuse by the industrialized world. But there's also hope and pride in collective efforts: the renewable energy transition, international agreements where nations come together, moments when humanity rises to meet an existential challenge.

Social media analyses reveal that this moral language—care, justice, global responsibility—forms a recognizable cluster distinct from contrarian or minimalist climate discourses (Falkenberg, M. et al, 2022). Like the economic liberty perspective, this isn't just a position on climate science. It's a comprehensive worldview about what matters, what's true, and what we owe to one another.

## 8.2.3 Two Tribes, Two Futures

What we see in these two perspectives isn't simply a disagreement over climate science or economic modeling. Each represents a coherent moral and epistemic framework with its own authorities, its own standards of evidence, its own emotional register, and its own vision of responsibility. One asks "What will happen to our prosperity and freedom?" The other asks "What will happen to vulnerable people and future generations?" Both questions matter—but they point toward very different answers about what we should do.

## 8.2.4 Summary Table – Climate Change

| Aspect | Economic Liberty TT (Skeptical/Minimalist) | Global Responsibility TT (Alarm/Activist) |
|---|---|---|
| Core Truth Bearer | "Strong climate policies will harm prosperity and freedom." | "Inadequate climate action is a grave moral wrong." |
| Truth Makers | GDP, jobs, energy prices; historical examples of regulatory cost. | Temperature trends, model projections, observed climate impacts. |
| Authorities | Pro-growth economists, industry, nationalist politicians. | Climate scientists, assessment bodies, environmental NGOs. |
| Trust Policies | Prioritize economic data; discount global institutions as politicized; short time horizon. | Prioritize scientific consensus and long horizons; weight justice claims heavily. |
| MAC Frame | Property, Group (nation), Fairness-as-Proportionality. | Kin (future generations), global Group, Reciprocity, Fairness-as-Equity. |
| EMFT Profile | Liberty/Oppression; Loyalty/Betrayal (nation vs. outsiders); Fairness-proportional. | Care/Harm; Fairness-equity; Authority; reinterpreted Liberty (freedom from climate harms). |
| Emotional Strategies | Fear of decline; pride in industry; resentment of elites. | Fear for vulnerable and future kin; guilt; hope in collective action. |

## 8.3   The AI Future

As artificial intelligence grows more powerful, two radically different perspectives have emerged about how humanity should respond. Each offers a coherent vision of what's at stake, what we should fear, and what we owe to the future (Waite 2023).

### 8.3.1  The Doomer Perspective

For those who hold the doomer view, the central truth is stark and terrifying: if advanced AI is developed without strict constraints, it is likely to cause human extinction or a catastrophe of comparable magnitude. This isn't casual pessimism—it's a conviction born from careful reasoning about what happens when intelligence vastly exceeds human capacity.

What validates this truth for them comes from several converging lines of argument. There are theoretical frameworks about superintelligence, instrumental convergence, and the control problem—the fundamental difficulty of ensuring that a system far smarter than us will do what we actually want rather than what we technically specified (Bostrom 2014; Russell 2019). There are thought experiments and simplified models showing how misaligned objectives can spiral into extreme outcomes. And increasingly, they point to early warning signs in current systems: emergent capabilities that weren't programmed in, opacity that makes systems impossible to fully understand, experiments revealing how easily AI systems can be misaligned from human intentions.

The moral logic here is about survival at the most fundamental level. This isn't about competing interests or trade-offs—it's about whether the human lineage continues to exist. As Curry, Mullins, and Whitehouse (2019) describe in their framework of morality as cooperation, this activates the deepest cooperative instinct: kin selection and group survival extended to the entire species.

Who do people in this camp trust? They turn to AI safety researchers and alignment theorists who have spent years thinking about these problems, to philosophers of existential risk who study civilizational-scale threats, and to technical insiders who have worked at the cutting edge and chosen to publicly warn about uncontrolled development. These are the voices, they believe, who understand what's actually being built and aren't blinded by commercial incentives.

This shapes how they evaluate information in distinctive ways. Low-probability, high-impact outcomes become decisive—when the stakes are literally everything, even a small chance of catastrophe demands action. They weight the testimony of technical experts and whistleblowers far more heavily than the reassurances of optimists who have financial stakes in rapid development. And they insist on a precautionary approach: those building powerful systems should bear a heavy burden of proof to demonstrate safety, not the other way around.

The emotional landscape of this worldview is dominated by care and harm considered at an existential scale—not just suffering, but the permanent elimination of everything humanity could become. There's sometimes a sense of oppression, of AI as a potential new overlord. And there's an element of purity, a concern about losing something uniquely and irreplaceably human (Haidt

2012). From the perspective of psychological research on decision-making, vivid catastrophic scenarios generate intense negative emotions that—without careful reflection—can be mistaken for evidence rather than recognized as emotional reactions (Kahneman 2011).

The emotional appeals that resonate include apocalyptic imagery and analogies: runaway paperclip maximizers that convert the entire Earth to their simple objective, fictional references like "Skynet," scenarios where humanity becomes irrelevant or extinct. There's moral urgency framed as duty: "We must slow or halt this until we understand it" becomes an obligation to future generations who won't exist if we get this wrong. And there's a valorization of whistleblowing and restraint as courageous virtues—speaking uncomfortable truths and choosing not to build something just because you can.

## 8.4 The Accelerationist Perspective

Those who embrace the accelerationist view start from a fundamentally different truth: rapid AI development is a moral imperative because it will dramatically improve human welfare. This isn't reckless optimism—it's a conviction that the greatest risk is moving too slowly while people suffer from problems AI could solve.

What makes this true for them comes from a different evidence base entirely. They point to current successes of AI: medical imaging systems that detect diseases earlier than human doctors, language models that democratize access to information and education, efficiency gains across industries that create abundance. They look at historical patterns showing how technological progress has consistently increased wealth and life expectancy (Simon J 1990). And they make arguments about how AI can help solve the most pressing global problems—disease, poverty, even climate change itself.

The moral logic here centers on cooperation and heroism at scale. This is about humanity and its innovators working together to build powerful tools that benefit everyone. As Curry, Mullins, and Whitehouse (2019) describe, this reflects mutualistic cooperation—the kind where everyone gains from collaborative achievement.

The authorities trusted by this group are tech entrepreneurs, engineers, and venture investors who are actually building these systems, futurist writers and techno-optimist thinkers who articulate positive visions of what technology can achieve (Andreessen 2023), and policy advocates who push for innovation-friendly regulation that doesn't strangle progress in its cradle. These voices, they believe, understand both what's possible and what's necessary to get there.

Their approach to information reflects these commitments. Past technological gains provide strong evidence about future benefits—the historical track record of innovation improving lives becomes a guide to what's coming. Concrete demonstrations of capability weigh more heavily than speculative risk forecasts that may never materialize. And crucially, they consider the opportunity cost of delay: every day we postpone deploying powerful AI is a day we lose potential benefits, lives that could have been saved, suffering that could have been prevented.

The emotional texture of this worldview also centers on care and harm—but the focus is on current and near-term suffering that AI might reduce. There's also a strong sense of fairness: that the eventual abundance created by AI will be broadly shared, lifting everyone. There may be deference to successful innovators as legitimate authorities who have proven their ability to create value. And there's an emphasis on liberty as the freedom to experiment, to build, to push boundaries (Haidt 2012).

The emotional appeals that resonate include inspiring visions of human flourishing: curing cancer, ending scarcity, exploring space, solving problems that have plagued humanity for millennia. There's pride in human ingenuity and a Promethean ethos—the conviction that it's humanity's nature and destiny to reach for greater power and capability. And there's a moral framing of caution as cowardice or costly delay: "Slowing AI costs lives we could have saved" becomes an indictment of excessive fear.

### 8.4.1 Two Tribes, Two Futures

The clash between these perspectives runs deeper than simple optimism versus pessimism. It's a fundamental disagreement about which cooperative challenge is primary: species survival under radical uncertainty, or rapid expansion of human capability through technological progress. It's about which moral and emotional reactions should guide our trust and our choices—fear of what we might create, or hope for what we might achieve.

As Bostrom (2014), Russell (2019), and Andreessen (2023) have each argued from their different positions, the stakes are genuinely enormous. But they disagree profoundly about which direction the greatest danger lies: in moving too fast without understanding what we're building, or in moving too slowly while people suffer from problems we have the tools to solve. Each perspective has its own internal coherence, its own authorities, its own standards of evidence, and its own vision of responsibility. The question is not which tribe is simply "right"—it's how we navigate between two legitimate but conflicting imperatives as we build the future.

### 8.4.2 Summary Table – AI Future

| Aspect | Doomer TT | Accelerationist TT |
|---|---|---|
| Core Truth Bearer | "Uncontrolled advanced AI is likely to destroy us." | "Rapid AI development will massively improve human welfare." |
| Truth Makers | Theoretical risk models, control-problem arguments, early warning signs (Bostrom 2014; Russell 2019). | Current AI successes, historical tech progress, projected benefits (Simon J 1990; Andreessen 2023). |
| Authorities | AI safety researchers, existential-risk philosophers, technical whistleblowers. | Tech entrepreneurs, engineers, techno-optimist writers. |
| Trust Policies | Precautionary, risk-averse; strong weight on expert alarms and worst-case scenarios. | Progress-oriented; strong weight on performance data and historical analogies; concern about opportunity costs. |
| MAC Frame | Kin/Group survival; existential cooperation problem. | Mutualism and Heroism; Promethean expansion of shared capability. |
| EMFT Profile | Care/Harm at an existential scale; Liberty/Oppression (fear of machine domination); some Purity (loss of "human essence"). | Care/Harm focused on problems AI can solve; Fairness (eventual abundance); Authority (innovators); Liberty (freedom to build). |
| Emotional Strategies | Fear, dread, urgency; valorization of restraint and whistleblowing. | Hope, excitement, pride; framing caution as harmful delay or timidity. |

## 8.5 Information Disorders as Conflicts of Narratives and Cooperative Frames

Information disorders—misinformation, disinformation, propaganda, and engineered echo chambers—can be understood within MEVIR 2 as attempts to reshape or exploit the trust process at multiple layers.

At the ontological level, disinformation often involves shifting the admissible truth-maker. When allegations of corruption are reframed as evidence that a political leader is fighting a shadowy "deep state," the narrative invites supporters to treat heroic Group cooperation and Deference to a charismatic leader as the relevant MAC domains, and Loyalty and Authority as the key EMFT foundations. Legal and evidential truth-makers such as court documents and financial records are downplayed as irrelevant artifacts produced by untrustworthy institutions.

At the moral level, propaganda works by constructing narratives that vividly activate certain foundations and suppress others. Ethnonationalist rhetoric, for example, heightens Loyalty and Purity by emphasizing threats from outsiders who purportedly contaminate the group's culture or gene pool. Once these foundations are activated, procedural reasoning is likely to search for evidence that vindicates group defense, and virtue failures such as confirmation bias and the bandwagon effect become more powerful. Public-health campaigns, even when well intentioned, may also function propagandistically if they rely primarily on fear appeals that activate Care and Purity without fostering transparent understanding (Johnson 2012).

Echo chambers can be modeled as trust lattices that have become self-referential. Trust policies are tuned to systematically reject any claim originating from an out-group authority, regardless of its factual content. Within such a lattice, belief revision is effectively disabled except for updates that reinforce existing anchors. Cooperative games and moral foundations are repeatedly framed in the same way, narrowing the agent's imagination about alternative frames. For instance, immigration may be presented exclusively as a Group and Property threat, never as a Reciprocity or global Kin concern.

Psychological operations (PSY-OPS) in military or political contexts can be described as deliberate efforts to reconfigure opponents' MAC domains and EMFT profiles. By amplifying internal divisions, a PSY-OP might encourage one subgroup to interpret a conflict primarily through Fairness-as-proportionality and Deference to local leaders, while encouraging another subgroup to view it through Liberty/Oppression and suspicion of those same leaders. The goal is to fragment cooperative games and undermine shared trust anchors.

MEVIR 2 does not in itself prevent such manipulations, but it provides a language for diagnosing them. By mapping which cooperative games are being invoked, which foundations are being triggered, and which authorities are being elevated or delegitimized, analysts can identify the strategic structure of information disorders rather than treating them as random noise.

# 9 Towards a MEVIR 2-Based Decision Support System

Although MEVIR 2 is primarily a descriptive framework, it suggests a design space for tools that assist individuals in making more conscientious and reflective trust decisions. A decision-support system based on MEVIR 2 would not tell users what to believe, but it would make their own trust processes visible and gently nudge them toward better epistemic practice.

When a user encounters a controversial article, the system might first infer which MAC domains and EMFT foundations the article's language is likely to activate. References to "our people," "traitors," and "defending the homeland" indicate a focus on Group cooperation and Loyalty; references to "protecting the vulnerable" and "future generations" indicate Kin and global Reciprocity with high Care. These inferences could be displayed in a way that draws attention to the moral framing. Seeing that a piece of content is primarily appealing to Purity and Authority, for example, may alert the user to the possibility of disgust-based or authoritarian manipulation.

As the user follows hyperlinks or performs searches, the system could map the emerging trust lattice: which claims are being accepted, which authorities are being relied upon, and which kinds of evidence are being ignored. It might detect that an immigration debate is being explored exclusively through sources emphasizing Group threat and Property concerns. An epistemic nudge could then suggest sources that frame the same issue through Reciprocity and Care, such as analyses of immigrants' economic contributions or reports on humanitarian conditions. The system would not force the user to consult these sources but would lower the cost of doing so.

Built-in diagnostics for common virtue failures could further enhance metacognition. If the system notices that the user repeatedly favors sources aligned with their prior beliefs and avoids those associated with out-groups, it might provide a gentle reminder about confirmation bias and the bandwagon effect. If the user appears to treat a charismatic figure as authoritative across many unrelated domains, the system might highlight the risk of the halo effect. The aim is not to shame the user but to provide them with vocabulary and visibility for their own patterns.

In health and climate domains specifically, a MEVIR 2-based system could help clarify cooperative frames. When a user resists a vaccine mandate on grounds of bodily autonomy, the system might present alternative framings that highlight Kin and Group reciprocity without dismissing Property and Liberty concerns outright. Similarly, when a user supports climate policies solely on the basis of global justice, the system might encourage engagement with arguments about economic Fairness-as-proportionality and national Group interests, thereby expanding their appreciation of the trade-offs at stake.

# 10 Conclusion and Directions for Future Work

MEVIR 2 has been presented as a virtue-informed, ontologically grounded, and morally hybrid framework for understanding how human beings form trust judgments in complex information environments. It integrates a procedural model of evidence elaboration, a virtue-theoretic account of the epistemic agent, and a hybrid MAC–EMFT moral model that links cooperative problems to intuitive moral responses.

Ontologically, MEVIR 2 insists that trust decisions are ultimately about judgments regarding truth-makers and admissible proxies. Agents do not simply accept sentences; they accept or reject implicit pictures of reality. Epistemically, the framework highlights the role of intellectual virtues and vices in governing how agents navigate between direct evaluation and deference to authority. Morally, MEVIR 2 explains how different emphases on cooperative domains and moral foundations can produce distinct but internally coherent trust lattices. These lattices anchor polarized debates in health, climate, and beyond.

The framework's main strength lies in its descriptive and diagnostic power. It makes explicit the structure behind disagreements that might otherwise appear as mere stubbornness or irrationality. It also suggests avenues for constructive engagement: moral reframing that respects but redirects foundational concerns; institutional reforms that diversify available authorities; educational programs that cultivate epistemic virtues and ontological literacy.

The framework reveals something both humbling and hopeful: we're not isolated rational agents choosing our beliefs through pure reason. We're tribal creatures, shaped by group-level selection pressures we barely perceive. But understanding these pressures gives us the possibility of building better tribes—moral-epistemic communities that serve both truth and cooperation, both individual flourishing and collective adaptation.

That's not a guarantee of success. Evolution is blind, and cultural evolution more so. But it's a path forward: recognizing that our epistemic lives are fundamentally communal, and that the challenge isn't to transcend our tribal nature but to evolve tribes worthy of our commitment.

Future work should aim to operationalize MEVIR 2 in empirical studies. Longitudinal research could track how individuals' MAC emphases, EMFT profiles, and trust lattices evolve across life events such as parenthood, major political shifts, or exposure to targeted information campaigns. Experimental work could test whether MEVIR-informed interventions—such as explicit cooperative reframings or virtue-based metacognitive prompts—actually improve the quality of trust decisions in domains like vaccination and climate policy.

Ultimately, MEVIR 2 suggests that any adequate theory of trust in the twenty-first century must engage simultaneously with logic, character, and cooperation. It is only by understanding how people think, who they are, and what kinds of social games they believe they are playing that we can hope to navigate the moral–epistemic turbulence of the current information age.

# References

Adamo, Greta, Anna Sperotto, Mattia Fumagalli, Alessandro Mosca, Tiago Prince Sales, and Giancarlo Guizzardi. 2024. "Unpacking the Semantics of Risk in Climate Change Discourses." In *Proceedings of the 14th International Conference on Formal Ontology in Information Systems (FOIS 2024)*. https://www.utwente.nl/en/eemcs/fois2024/resources/papers/adamo-et-al-unpacking-the-semantics-of-risk-in-climate-change-discourses.pdf

Alfano, Mark. 2012. "Expanding the Situationist Challenge to Responsibilist Virtue Epistemology." *The Philosophical Quarterly* 62 (247): 223–49 https://hdl.handle.net/20.500.14802/2888.

Amaral, G., V.d. Santos, E.H. Haeusler, G. Guizzardi, D. Schwabe, and S. Lifschitz. 2026. "Unpacking Trust: An Ontological Framework for Information Trustworthiness in Decision-Making." In Advances in Conceptual Modeling: ER 2025, edited by C. M. Fonseca, A. Bernasconi, S. de Cesare, L. Bellatreche, and O. Pastor. LNCS, vol 16190. Cham: Springer. https://doi.org/10.1007/978-3-032-08620-4_11.

Andreessen, Marc. 2023. "The Techno-Optimist Manifesto." Andreessen Horowitz. https://a16z.com/the-techno-optimist-manifesto/.

Bostrom, Nick. 2014. *Superintelligence: Paths, Dangers, Strategies*. Oxford: Oxford University Press. ISBN: 9780199678112

Bernasconi, Alessio, Giancarlo Guizzardi, Tiago Prince Sales, and others. 2022. "An Ontological Account of Viral Entities and Events." In *Proceedings of the 13th International Conference on Formal Ontology in Information Systems (FOIS 2022)*.

Curry, Oliver Scott. 2016. "Morality as Cooperation: A Problem-Centred Approach." In *The Evolution of Morality,* edited by Todd K. Shackelford and Ranald D. Hansen, 27–51. Berlin: Springer.

Curry, Oliver Scott, Daniel Austin Mullins, and Harvey Whitehouse. 2019. "Is It Good to Cooperate? Testing the Theory of Morality-as-Cooperation in 60 Societies." *Current Anthropology* 60 (1): 47–69.

David, Marian. 2025. "The Correspondence Theory of Truth." In *The Stanford Encyclopedia of Philosophy,* edited by Edward N. Zalta. https://plato.stanford.edu/archives/sum2025/entries/truth-correspondence/.

Falkenberg, Max, Alessandro Galeazzi, Maddalena Torricelli, Niccolò Di Marco, Francesca Larosa, Madalina Sas, Amin Mekacher, Warren Pearce, Fabiana Zollo, Walter Quattrociocchi, and Andrea Baronchelli. 2022. "Growing Polarization around Climate Change on Social Media." Nature Climate Change 12 (12): 1114–1121. https://doi.org/10.1038/s41558-022-01527-

Gilboa, Itzhak. 2009. *Theory of Decision under Uncertainty*. Cambridge: Cambridge University Press.

Guizzardi, Giancarlo, and Nicola Guarino. 2024. "Ontological Unpacking and the Analysis of Conceptual Models." *Applied Ontology* 19 (1): 1–28.

Guizzardi, Giancarlo, Tiago Prince Sales, and others. 2021. "Towards an Ontology of Viral Epidemiology." In *Proceedings of the 12th International Conference on Formal Ontology in Information Systems (FOIS 2021)*.

Haidt, Jonathan. 2012. The Righteous Mind: Why Good People Are Divided by Politics and Religion. New York: Pantheon.

Hancock, Jeffrey T., Emma S. Levine, David M. Markowitz, and others. 2023. "Deception, Trust, and Technology." *Annual Review of Psychology* 74: 573–99.

Hunter, Anthony, and Richard Booth. 2015. "Trust and Revision for Structured Beliefs." In Proceedings of the 24th International Joint Conference on Artificial Intelligence (IJCAI 2015).

Johnson, Bethany. 2012. "Fear Appeals and Public Health: The CDC's 'Tips from Former Smokers' Campaign." *Journal of Health Communication* 17 (9): 1081–93.

Kahneman, Daniel. 2011. *Thinking, Fast and Slow.* New York: Farrar, Straus and Giroux.

McNeace, Ryan, and Jeremy Sinn. 2018. "Schwartz's Theory of Basic Human Values and Political Ideology." *Personality and Individual Differences* 121: 116–21.

Pereira, Luis Moniz, Andrea Tettamanzi, and Serena Villata. 2011. "Changing One's Mind: Erase and Rewind." In *Logic Programming, Knowledge Representation, and Nonmonotonic Reasoning,* edited by Marcello Balduccini and Tran Cao Son, 14–26. Berlin: Springer.

Russell, Stuart. 2019. Human Compatible: Artificial Intelligence and the Problem of Control. New York: Viking.

Schertzer, Robert. 2025. "Climate Politics and the Crisis of Trust." *Global Environmental Politics* 25 (2): 55–78.

Schwabe, Daniel. "The MEVIR Framework: A Virtue-Informed Moral-Epistemic Model of Human Trust Decisions." *arXiv preprint arXiv:2512.02310* (2025).

Simon, Herbert A. 1990. "Bounded Rationality." In *Utility and Probability,* edited by John Eatwell, Murray Milgate, and Peter Newman, 15–18. New York: Palgrave Macmillan.

Simon, Julian L. 1990. *Population Matters: People, Resources, Environment, and Immigration*. New Brunswick, NJ: Transaction Publishers.

Sober, Elliott, and David Sloan Wilson. 1998. *Unto Others: The Evolution and Psychology of Unselfish Behavior*. Cambridge, MA: Harvard University Press.

Varin, Jordan. 2025. "Polarization and the Narrative Politics of Climate Change." *Environmental Politics* 34 (1): 91–112.

Waite, Thom (2023) - Doomer vs Accelerationist: the two tribes fighting for the future of AI | Dazed https://www.dazeddigital.com/life-culture/article/61411/1/doomer-vs-accelerationist-two-tribes-fighting-for-future-of-ai-openai-sam-altman

Wilson, David Sloan. 2002. *Darwin's Cathedral: Evolution, Religion, and the Nature of Society*. Chicago: University of Chicago Press.

Wu, Yuxin, Marcelo Arenas, and Sergio Gómez. 2017. "A Framework for Trust in Data and Information." *Journal of Web Semantics* 45: 1–15.

Zagzebski, Linda. 1996. Virtues of the Mind: An Inquiry into the Nature of Virtue and the Ethical Foundations of Knowledge. Cambridge: Cambridge University Press.

# Executive Summary

The core innovation of MEVIR 2 is the concept of **"Truth Tribes" (TTs)**—emergent communities that share aligned procedural, virtue-theoretic, and moral-epistemic profiles. These tribes construct internally coherent "trust lattices" that often remain mutually unintelligible to outsiders, explaining why polarization persists despite shared evidence. The framework treats cognitive biases not as random errors, but as systematic failures of **epistemic virtue**. By applying MEVIR 2, analysts can deconstruct narratives to reveal the hidden structures of trust, identifying why certain authorities or facts feel salient to different groups.

---

# 1.    Practical Application: Step-by-Step Analysis

To apply the MEVIR 2 framework to a given narrative (e.g., a speech, article, or social media thread), follow these four procedural levels.

## 1.1   Level 1: Ontological Unpacking

This level identifies what the narrative assumes to be real and what makes its claims "true".
- **Action:** Distinguish between **Truth Bearers** (the claims being made) and **Truth Makers** (the aspects of reality that confer truth on those claims).
- **Questions to Pose:**
    - What are the core "Truth Bearers" or claims in this narrative?
    - What specific "Truth Makers" are implicitly posited to make these claims true (e.g., statistical data vs. personal anecdotes)?
    - How would this claim be "unpacked" into its constituent objects, relations, and dispositions?
- **Evaluation Criteria:**
    - **Specificity:** Does the unpacking reveal a precise structure (e.g., viral load vs. state overreach)?
    - **Admissibility:** Are the posited truth-makers considered legitimate by the target audience or dismissed as "orthogonal" to the moral stakes?

## 1.2   Level 2: Procedural Trust Elaboration

This level maps the mechanical structure of evidence gathering and the "trust lattice" supporting the narrative[151515].
- **Action:** Trace the recursive search for evidence until it terminates at a **Trust Anchor**[16161616].
- **Questions to Pose:**
    - What is the "trust lattice" or network of supporting statements? [17]
    - What are the primary **Trust Anchors** (e.g., accepted authorities, pre-trusted beliefs, or resource exhaustion)? [18181818]
    - What **Trust Policies** (standards of evidence, source reliability rules) are being applied? [19]

- **Evaluation Criteria:**
  - **Consistency:** Does the lattice adhere to the principle of minimal change when facing conflicting info?
  - **Stickiness:** Is the computational cost of rejecting a central anchor (e.g., a core belief) too high for the agent to accept a correction?

## 1.3  Level 3: Virtue-Epistemic Assessment

This level characterizes the epistemic character of the agent or narrative producer
- **Action:** Evaluate whether the agent is operating as a **Conscientious Agent** or exhibiting **Epistemic Vices**.
- **Questions to Pose:**
  - Is the agent using **Path 1** (direct evaluation) or **Path 2** (deference to authority)?
  - Are intellectual virtues (humility, open-mindedness) or vices (arrogance, laziness) evident?
  - Is there evidence of **Epistemic Encapsulation** (following tribal rules while being severed from external reality)?
- **Evaluation Criteria:**
  - **Bias Identification:** Can observed deviations be mapped to specific failures, such as **Confirmation Bias** (corrupting Path 2 for validation) or **Overconfidence** (misapplying Path 1)?
  - **Character:** Is the agent's "felt sense of virtue" actually a rationalization of tribal needs? [28282828]

## 1.4  Level 4: Hybrid Moral Modeling (MAC–EMFT)

This level identifies the pre-rational cooperative problems and intuitive foundations shaping the trust decision.
- **Action:** Map the narrative to **Morality-as-Cooperation (MAC)** domains and **Extended Moral Foundations (EMFT)**.
- **Questions to Pose:**
  - Which **MAC Domain** is being invoked (e.g., Kin Selection, Mutualism, Reciprocity, Possession)?
  - Which **EMFT Foundation** is triggered (e.g., Care/Harm, Liberty/Oppression, Purity/Sanctity)?
  - What is the **Asymmetrical Attribution** (explaining the "in-group" via context and the "out-group" via bad character)?
- **Evaluation Criteria:**
  - **Cooperative Game:** Does the agent believe they are playing a "Hawk–Dove" contest or a "Public Goods" reciprocity game?
  - **Moral SALIENCE:** Which facts are noticed or ignored based on the "moral taste buds" being activated? [3]

---

## 1.5  Verification Checklist

1. **Procedural Model:** Confirmed definition of trust lattices and anchors[41].
2. **Virtue Model:** Confirmed Path 1 vs. Path 2 and Zagzebski's influence[42424242].
3. **Moral Model:** Verified MAC and EMFT integration and specific domain/foundation mappings.

4. **Truth Tribes:** Confirmed definition as emergent social groups aligned on all three pillars[44].
5. **Biases:** Verified reinterpretation of common biases as virtue failures.

# 2 Case Study

The following analysis applies the **MEVIR 2 framework** to the current debate between the capabilities of **Generative AI** (GenAI) and the pursuit of **Artificial General Intelligence** (AGI). This discourse is characterized by a "Truth Tribe" schism between those who view modern models as a direct path to human-level intelligence and those who see them as "masterful parrots" lacking true causal understanding.

The debate over AI's evolution is not merely technical; it is a **moral-epistemic conflict** shaped by divergent "Truth Makers" and tribal loyalties. While **Generative AI** focuses on content creation through pattern mimicry, **AGI** aims for broad, human-like cognitive autonomy across all domains. The MEVIR 2 framework reveals that "Doomer" and "Accelerationist" tribes are not just disagreeing on facts, but are operating in entirely different **moral-epistemic worlds** with distinct trust anchors and virtue failure modes.

---

## MEVIR 2 Analysis: The AI Evolution Debate

## 2.1 Level 1: Ontological Unpacking

This level identifies what specific aspects of reality are believed to "make" AI claims true.

- **Truth Bearer:** "Scaling current generative models is the path to AGI."
- **Truth Makers (Accelerationist):** Compute power, dataset size (45TB+), and benchmark scores like MMLU (78%+).
- **Truth Makers (Skeptic/Doomer):** Causal understanding, sensorimotor learning (embodied cognition), and "g-factor" general intelligence.
- **Evaluation Criteria:** Does the narrative treat "fluent text" as evidence of "thinking" (mimicry) or does it demand "world-modeling" (causality)?

## 2.2 Level 2: Procedural Trust Elaboration

Mapping the "trust lattice" and determining where the search for evidence terminates.

- **Trust Anchors:** Accelerationists often anchor in **Accepted Authorities** like industry CEOs (e.g., Sam Altman) or "techno-optimist" manifestos. Doomers anchor in independent safety researchers and alignment theorists (e.g., Nick Bostrom, Eliezer Yudkowsky).
- **Trust Policies:** Accelerationists prioritize "empirical success" and "opportunity costs" of delay. Doomers prioritize "precautionary principles" and "low-probability/high-impact" extinction risks.
- **Evaluation Criteria:** Is the trust lattice anchored in corporate progress reports or in theoretical safety proofs?

## 2.3 Level 3: Virtue-Epistemic Assessment

Characterizing the intellectual character of the participants.

- **Path 1 (Direct Reliance):** Users often trust their "vibe" that an AI is conscious because it is persuasive—a failure of **attentiveness** to the underlying math.
- **Vices Incurred:** * **Halo Effect:** Granting technical authority to charismatic founders on matters of global ethics.
    - **Overconfidence/Dunning-Kruger:** Believing one understands the "black box" of LLMs because the output is human-like.
- **Evaluation Criteria:** Is the agent exercising **epistemic humility** regarding the "opacity" of these models?

## 2.4 Level 4: Hybrid Moral Modeling (MAC–EMFT)

Identifying the cooperative games and intuitive "moral taste buds" being triggered.

- **MAC Domains:**
    - **Heroism (Accelerationist):** AI as the tool to solve cancer, climate change, and poverty.
    - **Group Survival (Doomer):** AI as an existential "Hawk" that may out-compete the human "Dove".
- **EMFT Foundations:**
    - **Liberty:** Freedom to innovate without government "strangleholds".
    - **Care/Harm:** Preventing the "ultimate harm" (extinction) or the "current harm" (disease AI could cure).
    - **Purity:** Concern about losing "human essence" to machines.
- **Asymmetrical Attribution:** Accelerationists label critics as "luddites" (dispositional flaw); Doomers label developers as "reckless/greedy" (dispositional flaw).

---

## 2.5 Characterizing the Resulting Truth Tribes (TTs)

### 2.5.1 The "e/acc" (Effective Accelerationist) Tribe

- **Belief Structure:** Progress is a moral imperative; intelligence is a cosmic good.
- **Epistemic Authorities:** Tech entrepreneurs and "frontier" lab results.
- **Moral Value:** Heroism and Mutualism (global prosperity).

### 2.5.2 The "AI Safety/Doomer" Tribe

- **Belief Structure:** Advanced AI is an uncontrollable agent; P(doom) is non-zero.
- **Epistemic Authorities:** Alignment researchers and existential-risk philosophers.
- **Moral Value:** Care/Harm (existential scale) and Group Survival.

---

## 2.6 Verification Checklist

1. **Ontological Unpacking:** Distinguished between GenAI (content) and AGI (cognition).
2. **Procedural Model:** Identified CEOs vs. Safety Researchers as conflicting trust anchors.
3. **Virtue Model:** Mapped the "Halo Effect" and "Overconfidence" to the AI discourse.
4. **Moral Model:** Applied MAC (Heroism vs. Survival) and EMFT (Liberty vs. Care) to tribal narratives.
5. **TT Identification:** Explicitly characterized the "e/acc" and "Doomer" Truth Tribes.

| Layer | Doomer TT | Accelerationist TT |
|---|---|---|
| **Ontological** | **Truth Makers:** Theoretical risk models and control-problem logic. | **Truth Makers:** Current successes (GenAI) and historical tech patterns. |
| **Procedural** | **Anchors:** AI safety researchers and alignment theorists. | **Anchors:** Tech entrepreneurs and engineers. |
| **Moral (MAC)** | **Games:** Kin/Group Survival (existential threat). | **Games:** Mutualism and Heroism (Promethean expansion). |
| **Virtue/Bias** | **Risk:** High-arousal fear imagery acting as a proxy for evidence. | **Risk: Halo Effect**—granting technical CEOs authority over global ethics. |